C:\SLC-103P\SLC103PV15.doc /20100819

ACCEPTED FOR PUBLICATION IN
PLANETARY AND SPACE SCIENCE
2010

**"Secular Light Curve of Comet**

**103P/Hartley 2, target**

**of the EPOXI Mission".**

Ignacio Ferrín,
Center for Fundamental Physics,
University of the Andes,
Mérida 5101, VENEZUELA
ferrin@ula.ve

Number of pages: 30+11= 41

Number of Figures: 11

Number of Tables: 2

Proposed Running Head:

**"Secular Light Curve of Comet 103P"**

Name and address for editorial correspondence:

Dr. Ignacio Ferrín,
Apartado 700,
Mérida 5101-A,
Venezuela,
South America.

email address:

ferrin@ula.ve

**Abstract**

In support of the EPOXI mission, we have compiled and updated the secular light curve (SLC) of comet 103P/Hartley 2, a Jupiter Family comet, the next target of the Deep Impact extended mission. We have collected 845 observations from 1997, 121 from 1991 and 25 from 2004, and have added our own observations in 2005-2006. The main results of this investigation are: **a)** Of the order of 30 photometric parameters are measured and listed in the plots of this comet, over 20 of them new. Paper I (2005, Icarus 178, 493-516) gives the definitions of the parameters measured from the SLCs, although a brief description is included here. The turn on point of activity is -4.2±0.1 AU from the Sun, which corresponds to -400±40 d before perihelion. The total active time is $T_{ACTIVE}$ 1484 ±43 d. For comparison the active time of comets 1P/Halley and 9P/Tempel 1 are $T_{ACTIVE}$ = 1992 and 1069 days. **b)** 103P is a young dwarf comet, young because it has a photometric age P-AGE(1,1) = 15±2 cy (comet years), and dwarf because its diameter is $D_{EFFE}$ = 1.14±0.16 km. For comparison comets 1P/Halley and 9P/Tempel 1 have photometric ages P-AGE= 7.1 cy and P-AGE= 22 cy (Paper V) and diameters 9.8 km and 5.5 km. **c)** The nucleus is *very active* as can be deduced from the fact that the amplitude of the SLC is $A_{SEC}$= 10.8±0.1 mag in 1997. For comparison comets 1P and 9P have $A_{SEC}$ = 10.8 and 9.0 mag. Thus this comet is as active as 1P/Halley, but much smaller. **d)** This comet belongs to the class of *spill-over* comets defined in Paper VI. After remaining active up to aphelion, the comet spills-over its activity into the next orbit. **e)** 103P also belongs to the family of *comets that exhibit a break point* in their SLC. Two linear laws change slope at a break point located at $R_{BP}$ = -1.20±0.1 AU before perihelion, and magnitude $m_{BP}$ = 9.4±0.1. The slope of the first linear law after turn on is n= +9.44. This value can be compared with the slope of comets Hale-Bopp, 1P/Halley and 9P/Tempel 1: +10.3, +8.92 and +7.7 respectively. Since the SLC after turn on does not exhibit curvature in any of these comets, it is safe to conclude that sublimation is not controlled by water ice. The controlling substance may be CO or $CO_2$ ice. $CO_2$ has been detected spectroscopically. **f)** The water budget of this comet is calculated. The comet expends 1.88 $10^{10}$ kg of water per apparition vs 5.5 $10^{11}$ kg for comet 1P/Halley and 1.28 $10^{10}$ kg for comet 9P/Tempel 1. A new water-budget age is defined and it is found that WB-AGE = 19 cy vs WB-AGE = 0.65 cy for 1P/Halley and 28 cy for 9P. **g)** The values of Afρ are calculated on two dates and we find dust production rates of 106 and 37 kg/s, decreasing with solar distance as expected.

127

128

**h)** The thickness of the layer lost per apparition is calculated, and we find $\Delta r = 39$ m where r is the radius. Since the radius of this comet is r = 570 m, it is found that at the present rate the comet may disappear in only 17 revolutions (~109 y). **i)** By chance all comets visited by spacecraft have small photometric ages and thus are young objects. 103P follows the same trend. Thus it is expected that the surface morphology that will be found in future EPOXI images should be very similar to that of previous comets. It is suggested that any new mission to a comet should be made to a methuselah object (P-AGE > 100 cy) and several are proposed.

137

138

139

140

141

142

143

144

145

146

147

148

149

150

151

152

153

154

155

156

157

158

159

160

## 1. Introduction

In support of the EPOXI mission (Extra-solar Planet Observation and eXtended Investigation), we have compiled and updated the secular light curve of the Jupiter family comet 103P/Hartley 2. EPOXI is the continuation of the Deep Impact mission (A'Hearn and Combi, 2007, and papers in that issue) that successfully observed the collision of an impactor with the nucleus of comet 9P/Tempel 1 (A'Hearn et al. 2005, and papers therein).

The results presented in this work parallel those of comet 9P/Tempel 1 (Ferrín, 2007), thus the same parameters are derived. Since 9P/Tempel 1 was studied in depth, it may be interesting to compare the present results with the former one, to place this comet in perspective, and to predict the value of some parameters. We will present the results with a minimum of interpretation.

In Paper I (Ferrín, 2005a), we have introduced the concept of secular light curve of a comet (SLC), which includes two views, the log plot and the time plot. The SLCs provide a wealth of information on the photometric and physical properties of comets as can be deduced from subsequent works (Papers II, III, IV, V, VI, 2005b, 2006, 2007, 2008, 2010), that could serve to complement other studies. In particular the *Atlas of Secular Light Curves of Comets* (Paper VI) is useful to get acquainted with the SLC curves, to see the rich variety of shapes exhibited by 27 comets, and to place into perspective the results presented in this paper.

Two plots are needed to characterize these objects because they provide information on different physical parameters in two different phase spaces. The importance of the time plot is that time runs uniformly in the horizontal axis, thus showing the brightness history of the object. Additionally the time plot can be converted to water production rate, allowing the calculation of the water budget of the comet (defined in Section 4). The importance of the log R plot is that power laws in R ($R^n$, where R= sun-comet distance), plot as straight lines, allowing the calculation of the turn on and turn off of activity. The log R plot is reflected at R=1 AU, to allow the determination of the absolute magnitude by extrapolation, thus the designation *reflected double-log R plot.*

195    To produce the SLCs we had to calculate ephemerides, absolute
196  magnitudes, separate nuclear observations, etc.   A key to the photometric parameters is
197  given in Paper I.  An updated key can be found in Paper VI.   A short description is given
198  next.

199

200    The magnitude at $\Delta$, R, $\alpha$, is denoted by V($\Delta$, R, $\alpha$), where $\Delta$ is the comet-Earth
201  distance, R is the Sun-comet distance, and $\alpha$ is the phase angle.  Briefly the parameters
202  listed in the log R plot are: q, the perihelion distance in AU, Q the aphelion distance also in
203  AU, Log Q, to know the extension of the plot, the turn on point $R_{ON}$, the turn off point $R_{OFF}$,
204  the asymmetry parameter $R_{OFF}/R_{ON}$, the magnitude at turn on $V_{ON}(1,R,0)$, the magnitude
205  at turn off $V_{OFF}(1,R,0)$, the absolute magnitude before perihelion $m_1(1,-1)$, the absolute
206  magnitude after perihelion $m_1(1,+1)$, the mean value of both $<m_1(1,1)>$, the value at LAG
207  (see definition below),  $m_1(1,LAG)$, the slope of the SLC just after turn on, the slope just
208  before perihelion, the slope just after perihelion, the absolute nuclear magnitude $V_N(1,1,0)$,
209  the amplitude of the secular light curve $A_{SEC}(1,1)= V_N(1,1,0) - <m_1(1,1)>$, the effective
210  diameter $D_{EFFE}$, the photometric age P-AGE(1,1) in comet years (cy), and the photometric
211  age calculated at LAG, P-AGE(1,LAG).   In the top line of the plot, right hand side, JF13/1
212  means that this is a Jupiter family comet of 13 comet years of age 1 km in diameter.  V.09
213  is the version of the plot, and Epoch indicates the year of the observations that best define
214  the envelope of the light curve.

215

216    The photometric age and the time age are defined in Papers I and II and are an
217  attempt to define the age of a comet through its activity:

218

219    P-AGE (1,1) = 1440 / [ $A_{SEC}$ (1,1) * (-$R_{ON}$ + $R_{OFF}$ )]  comet years (cy)          (1)
220

221    T-AGE (1,1) =  90240 / [ $A_{SEC}$ (1,1) * (-$T_{ON}$ + $T_{OFF}$ )]  comet years (cy)          (2)
222

223  where $A_{SEC}$ (1,1) = $V_N(1,1,0)$ – <m(1,1)>.   Thus $A_{SEC}$ (1,1), P-AGE(1,1) and T-AGE(1,1)
224  are proxies for activity and thus age and are calculated at $\Delta$ = 1 AU and R = 1 AU to be
225  able to compare comets with different q.   The validation of the concept of photometric
226  age, P-AGE, as a measure for age, is studied in Section 5.  P-AGE has been scaled to
227  human years.

228

229        The time plot contains a listing of the perihelia plotted, the orbital period of the

230    comet, $P_{ORB}$, the next perihelion in format YYYYMMDD, the number of observations used

231    in the plot, $N_{OBS}$, the time lag at perihelion, LAG, the turn on time, $T_{ON}$, in days, the turn off

232    time, $T_{OFF}$ in days, the asymmetry parameter, $T_{OFF}/T_{ON}$, the active time, $T_{ACTIVE}$, the time-

233    age T-AGE(1,1) in comet years (cy), the slope of the SLC at turn on, $S_{ON}$ .

234

235        *The importance of these secular light curves is that they show not only what we*

236    *know, but also what we do not know, thus pointing the way to meaningful observations.*

237    Figures 4 and 5 show the regions where observations are missing and thus are desirable

238    in the next apparition.  In particular observations after aphelion would determine the extent

239    of activity and thus $R_{OFF}$.   It is interesting to notice that the SLCs have to be done for each

240    apparition.  Since the comet evolves, after many apparitions it will be possible to construct

241    a movie of the comet in two phase spaces.  *Thus each SLC is a frame of a movie.*

242

243        In this paper we have collected 840 observations in 1997, 112 in 1991 and 25 in

244    2004 gathered from the literature, to construct the SLCs.   We will find that this Jupiter

245    family (JF) comet belongs to the family of *spill over comets*, those whose activity spills

246    over from one orbit to the next and to the family of *break point comets,* those  that exhibit

247    a break in the power law before perihelion (Paper I).     Additionally the information

248    presented in this study fills the important gap of photometric information and will be useful

249    to plan and interpret future observations of this comet.

250

251    **2. Photometric system**

252        The photometric system used in this work is described in Paper VI but a short

253    description follows.

254

255    *Envelope*   No corrections were applied to the observations.  The methodology outlined in

256    that paper makes no corrections and lets the brightest observations define the envelope.

257    It was found and justified there, that the *envelope* of the observations is the best

258    representation of the SLC of comets.  The justification is that all visual observations of

259    comets are affected by several effects, all of which decrease the perceived brightness of

260    the object by washing out the outer regions of the coma: moonlight, twilight, atmospheric

261    absorption (low altitude), haze, cirrus clouds, dirty optics (dust on mirror), excess

262    magnification, large aperture, lack of collimation, excess f/ number, lack of dark

263  adaptation, the delta effect, insufficient imaging exposure, insufficient measuring
264  aperture, frame of view smaller than coma, etc. *There are no corresponding physical*
265  *effects that could increase the perceived brightness of a comet.*  Thus the envelope is the
266  correct interpretation of the SLC and it does not require corrections of any kind.   This rule
267  is confirmed observationally.  The top observations in Figure 4 show a sharp edge, while
268  the bottom observations have no edge and are distributed in a diffuse area.

269

270      To define the envelope we follow the same procedure of Sosa and Fernandez
271  (2009).  We define 10-20 days time bins (or $\Delta$Log R = 0.1-0.2 in the log plot), and select
272  data points in the 5% upper percentile.  These selected data points are fitted by least
273  squares with straight lines or polynomials of up to degree 4.   The fitted lines define the
274  envelope.

275

276  *Infinite aperture magnitudes*     The comet exhibited at all times a prominent coma, so no
277  nuclear magnitude could be derived.   Instead, *infinite aperture magnitudes* had to be
278  measured (Paper II).

279

280      It has been known for some time (Green, 1997) that there is a lack of flux in many
281  CCD observations.   The reasons for this lack of flux are not always clear.   Fink et al.
282  (1999) have considered this problem and have an interesting suggestion:   *"It is rather well*
283  *known that visual observations in general, are several magnitudes brighter than CCD*
284  *measurements and reconciling CCD visual magnitudes is "more of an art than a science".*
285  *Figure 4* [of their work] *shows why this is the case.  As we increase the aperture size on*
286  *our CCD images we get brighter magnitudes.   It is evident by the progression of*
287  *magnitudes for increasing aperture size that if we integrate farther out we would readily*
288  *reach the magnitudes reported by visual observers.  The eye is essentially a logarithmic*
289  *detector and has a larger dynamic range than a CCD so that it effectively includes more of*
290  *the coma than an individual non-saturated CCD exposure".*

291

292      To avoid this error we have to measure all the light from the comet.   The problem
293  can be illustrated looking at Figure 1 where we show an image of comet 103P/Hartley 2
294  with a normal stretching of the image, and with a forced stretching.    Since large
295  photometric apertures have to be used to extract the whole flux, the image has previously
296  been cleaned of nearby stars, using a background cloning tool.   Stretching refers to the 297  maximum and minimum pixel intensity to display the image in the computer monitor.
298  The normal stretching is selected by the computer, who does not know about preserving
299  the flux.   The computer monitor displays a deceivingly faint image.  It turns out that a
300  forced stretching reveals a much larger coma than expected (Figure 1b).  Figure 1 reveals
301  that it is all too common to use photometric CCD apertures that do not extract the whole
302  flux from the comet.   This has been called the *insufficient CCD aperture error* (Ferrín,
303  2005b), and manifest itself by producing measurements that lie much below the envelope
304  of the observations (see Figures 4, 5, 6).   Of course the optimal extraction aperture
305  depends on the brightness and distance of the comet.

306

307       In Figure 2 we show the method we used to extract an infinite aperture magnitude,
308  IAM, from now on denoted as $R_{IAM}$ , $V_{IAM}$, the R and V magnitudes, respectively.   The
309  flux is measured with increasing CCD photometric apertures, and *the asymptotic value at*
310  *infinite is the infinite aperture magnitude.*  Notice how the calibration star #14 converges
311  very rapidly to an infinite aperture magnitude.  Not so the comet, that requires an aperture
312  with more than 90 pixels in diameter to extract a total flux.   Since the scale of this
313  telescope is ~1"/pixel, 90 pixels represent 93" of diameter or 1.55' of diameter, clearly
314  much larger than the usual apertures published in the literature to extract magnitudes.

315

316       Infinite aperture magnitudes should be adopted whenever the comet exhibits a
317  coma.   Currently we know only of one amateur group that carries out this type of
318  observation  on  a  systematic  way.    They  are  the  Spanish  group  at
319  http://www.astrosurf.com/cometas-obs.   They present tables of the magnitudes of comets
320  measured with increasing apertures.  This information is enough to derive an IAM.  Their
321  work should be continued and expanded.   Others groups would do well in adopting their
322  methods of observation.

323

324  **3. Comet 103P/Hartley 2**

325

326  *Observations*   Comet 103P/Hartley 2 was observed by the author on five nights with the
327  1 m f/3 Schmidt telescope of the National Observatory of Venezuela over a period of over
328  two years.  The Schmidt is equipped with a mosaic camera of 4x4 CCDs of 2048x2048
329  pixels each, for a grand total of 64 Megapixels, working in the drift scan mode.  No filter
330  was used to increase the signal to noise ratio.  The maximum response curve of the CCD

331 between 6500 and 7500 A, is approximated    fairly well by a broad band R filter.  The log
332 of the observations is in Table 1.

333

334      On January 12, 2005, 12 exposures of 142 s each were made with the 1 m Schmidt
335 telescope.  On the same night 6 exposures were obtained with the 1 m reflector, nearly
336 simultaneously.  The 1m reflector is equipped with a Tektronic CCD of 2048x2048 pixels.
337 On January 14, 2005, 6 exposures of 300 s each were made with the 1 m reflector.  The
338 comet was re-observed on March 1$^{st}$, 2006 with the 1 m Schmidt telescope, for 700 s
339 when it was at R= +5.03 AU post-perihelion and found to be fainter but still active.   The
340 comet was targeted for the last time on February 16 and 20$^{th}$ of 2007.  On the first date 6
341 images were combined to produce an exposure of 834 s.  On the second date 13 images
342 were combined for a total exposure of 1807 s.   The comet could not be detected.  We
343 determined very carefully the limiting magnitude and we found m(limit)= 20.9 for the first
344 date and 21.3 for the second.   These upper limits are not sufficient to constrain the SLC
345 beyond R= +5.78 AU, $\Delta$t= +1004 d.  Deeper images are needed at these distances.   The
346 nights were photometric.  The images are shown in Figure 3.

347

348      Processing was done using standard reduction procedures.   Calibration of the
349 magnitudes was done using the R-band magnitudes of the USNO-SA2.0 catalog.   To
350 avoid errors from poor magnitude stars, a least squares calibration of the field was done
351 using no less than 15 stars.  Stars deviating more than 3$\sigma$ from the linear fit were dropped.
352 The magnitudes derived on January 12$^{th}$ of 2005 from the Schmidt and the reflector using
353 a different focal ratio (f/3 vs f/21) differ by about +0.4 magnitudes, implying that the 21 m
354 focal distance looses flux.  It has been known for a long time that large f/ ratios loose flux
355 (Green, 1997).  Using the mean color index < V-R > = 0.41±0.16 from Table 1 of Paper II,
356 the R magnitudes were converted to visual V and are listed in Table 1.   These values are
357 plotted in the SLC, Figure 4.

358

359 *Data sets*  Observations of this comet have been collected from several sources.  There
360 are numerous observations in the International Astronomical Union Circulars (IAUC 5304,
361 5312, 5324, 5340, 5346, 5359, 5372, 5391, 6747, 6791, 6808, 6820).   Many observations
362 come from the International Comet Quarterly (Green, 2006).    The Spanish group
363 mentioned above contributed IAM observations. Additionally the following researchers
364 have made observations of this comet (shown in Figure 4), Chen and Jewitt (1994) = CJ,

365 Licandro et al. (2000)= LTLRH, Lowry and Fitzsimmons (2001)= LF, Lowry et al.
366 (2003)= LFC, Groussin et al. (2004)= GLJT, Snodgrass et al. (2006)= SLF, Snodgrass et
367 al. (2010)= SMH.

368

369 Lowry and Fitzsimmons (2001) = LF observed the comet at R= +3.63 AU. Lowry et
370 al. (2003) = LFC, have observations at R= +4.57 AU. Licandro et al. (2000) = LTLRH,
371 observed at R= +4.73 AU. The largest detection distance of this author was R= 5.03 AU.
372 Chen and Jewitt (1994) = CJ, have observations of this comet at R= +5.12 AU. In all
373 those cases the comet was active. The agreement of these data sets is fairly good as can
374 be deduced from Figures 4 and 5. But the record distance was by Snodgrass et al. (2010)
375 at R = -5.492 AU before perihelion when the comet was inactive. The observation by
376 Snodgrass et al. (2010) made -855 days before perihelion, was critical to constrain the
377 region of activity of the comet, and is plotted in Figures 4 and 5.

378

379 *Errors* The visual observations are affected by the insufficient flux error mentioned in
380 Section 2. Many observations lie below the envelope so only the upper 5% is used to
381 define the envelope as defined in Section 2.

382

383 When this series of papers was initiated, we expended many hours trying to define
384 the envelope of the data in a mathematical way. Several algorithms were explored,
385 including some that define edges. The envelope was also fitted independently by several
386 students. The different fits did not differ by more than ±0.12 magnitudes. We finally
387 settled on the procedure followed by Sosa and Fernandez (2009) and described at the
388 beginning of Section 2 which can define the envelope to ±0.1 magnitudes. All the fits
389 were made by least squares.

390

391 The professional observations have errors of 0.05 to 0.15 magnitudes. The
392 observation of 103P on July 28[th], 2008 by Snodgrass et al. (2008), has an error of ±0.20
393 magnitudes, but it was made at the level of magnitude 26.6, thus it is very reasonable.

394

395 In general observational errors were reasonable and comparable to those of other
396 comets measured in this series of papers (confirmation papers I-VI).

397

398   *Nuclear and faint observations*   Nuclear dimensions have been determined by

399   Jorda et al. (2000) who used ISOCAM of ISO, and obtained D= 1.16±0.2 km.  Groussin et

400   al. (2004) = GLJT, revised that value to D= 1.42±0.26 km.    Lisse et al. (2009) found

401   $D_{NUC}$= 1.14 ± 0.16 km.    The value by Groussin et al. (2004) was obtained +46 d after

402   perihelion, and thus was highly contaminated by a coma.    In spite of this they did a

403   remarkable job of subtracting the coma. The value cited by Lisse et al. (2009) was

404   obtained at -5.5 AU from the Sun, and thus the coma was much less prominent, and the

405   diameter more near to the real value.   Thus we adopt $D_{EFFE}$ = 1.14 ± 0.16 km.  Using the

406   classification proposed in Paper VI (*The Atlas*), we conclude that this is a *dwarf comet* (D

407   < 1.5 km).

408

409         The absolute nuclear magnitude in the visual, $V_N(1,1,0)$, is related to the effective

410   diameter, $D_{EFFE}$, by a compact and amicable formula derived in Paper III:

411

412               $Log [ p_V D^2_{EFFE} / 4 ] = 5.654 - 0.4 V_N(1,1,0)$

413

414   where $p_V$ is the geometric albedo in the visual.    For comets for which the geometric

415   albedo has not been measured, it is common to adopt $p_V$ = 0.04.   Thus the previous

416   equation can be simplified even further:

417

418               $Log [ D^2_{EFFE} ] = 7.654 - 0.4 V_N(1,1,0)$                          (3)

419

420   which is very easy to remember.   Assuming a geometric albedo $p_V$ = 0.04, then the

421   absolute nuclear magnitude is $V_N(1,1,0)$= 18.85±0.32.

422

423   *LAG and absolute magnitude parameters in 1997*    The best observed apparition has

424   been that of 1997 with 840 observations.    Thus most photometric parameters will be

425   deduced from that apparition.   The comet exhibits a lag between the maximum brightness

426   and perihelion, LAG= +16±2 days, that may be a thermal lag or due to the pole pointing

427   toward the sun at that time, or a combination of both (Figure 5).

428

429         The absolute magnitude <m(1,1)> = 8.1±0.1.   The maximum magnitude at the LAG

430   value is m(1,LAG) = 7.7±0.1.

431

432 *Turn on, turn off distances, and break point.*    Figures 4 and 5 allow the determination of
433 the turn on distance. We find $R_{ON}$ = -4.2±0.1 AU, $R_{OFF}$ = -6.0±0.2 AU and $T_{ON}$ = -400±40
434 d and $T_{OFF}$ = -830±15 d.    Thus $R_{SUM}$ = 10.5±0.2 AU and $T_{ACTIVE}$= 1904±43 d.

435

436     For comparison comets 1P/Halley and 9P/Tempel 1 turned on at $R_{ON}$  = -6.15  and
437 -3.5 AU respectively, while their $R_{SUM}$ was $R_{SUM}$= 18.7 AU and 7.7 AU.  Their active time
438 was $T_{ACTIVE}$= 1992 d and 1070 d.   We see that 103P is intermediate between the two.

439

440     This comet also belongs to the class of comets that exhibit a break point in their
441 SLC.  Other members of this class are C/1995 O1 Hale-Bopp, 1P/Halley, 81P/Wild 2,
442 21P/Giacobini-Zinner, and 9P/Tempel 1 (Papers I-VI).  Two linear laws are needed to
443 describe the behavior of this comet after turn on, and they change slope at a break point
444 located at $R_{BP}$ = -1.20±0.1 AU before perihelion, and magnitude $m_{BP}$ = 9.4±0.1.  The slope
445 of the first linear law is n= +9.44±0.05 while the second linear law has n= +5.70±0.04.
446 The first value can be compared with the corresponding values of comets Hale-Bopp,
447 1P/Halley and 9P/Tempel 1, +10.3, +8.9 and +7.7 respectively.

448

449     In paper VI it is shown that the preferred explanation for the break point, is the
450 change from sublimating something more volatile than $H_2O$, to sublimating $H_2O$.

451

452 *Composition.*    Delsemme (1983) has a plot of production rate of several molecules, $H_2O$,
453 $CO_2$ and CO vs heliocentric distance.  The water ice curve shows curvature at 2 AU but
454 CO or $CO_2$ are in a linear regime.  Since the SLC after turn on does not exhibit curvature
455 in any of the above comets (1P, 9P, HB, 103P), it is safe to conclude that sublimation is
456 not controlled by water ice (up to the break point).  For comet 103P the controlling
457 substance at turn on may be CO or $CO_2$ ice.   $CO_2$ has been detected spectroscopically
458 (Weaver et al. 1994) finding a $CO_2/H_2O$ ratio ~4%.  The spectrum was observed by
459 ISOPHOT, SWS and ISOCAM, and $H_2O$ and $CO_2$ were detected, but not CO (Colangeli et
460 al., 1999; Crovisier et al., 1999, 2000).  Thus it seems safe to conclude that the large
461 slope after turn on is due to $CO_2$.

462

463     Thus the shape of the SLC at turn on gives some idea of the substance that
464 controls sublimation.

465

466    *Photometric-age and time-age*    From   Figure 4 we measure an amplitude of the
467    secular light curve $A_{SEC}$ (1,1) = 10.8±0.1 in 1997.  For comparison comets 1P/Halley and
468    9P/Tempel 1 had $A_{SEC}$ = 10.8 and 9.0.  From a sample of 27 comets (Paper VI), it is
469    apparent that there is a maximum observed value of  $A_{SEC}$ = 11.55±0.1 mag.  103P is
470    close to that maximum value, implying that we are seeing a relatively young object.

471

472         Then using $R_{SUM}$ = 10.5 AU and Equation (1) we find a photometric-age P-AGE
473    (1,1) = 13±2 cy, a relatively young object.   From Figure 5 and Equation (2), we find T-
474    AGE = 5±1 cy.

475

476         In Paper VI comets have been classified by size and age.  Smaller than 1.5 km in
477    diameter are dwarf, while those with 3 cy < P-AGE(1,1) < 30 cy, are young.  *According to*
478    *this classification 103P is a young dwarf comet.*

479

480    *Activity post-aphelion*    Above we found the turn off point located at $R_{OFF}$ = -5.4±0.1 AU.
481    Since the aphelion distance of this comet is Q= 5.86 AU the turn off point is after aphelion.
482    Figures 4 and 5 show that there is activity *after* aphelion.  Thus this comet belongs to the
483    group of *spill-over comets*, those whose activity spills over from one orbit to the next.
484    Alternatively spill-over comets can be defined as *"those whose magnitude do not reach to*
485    *nuclear when the solar distance reaches to aphelion".*

486

487         Other probable members of this group (see Paper I) are 81P/Wild 2, 19P/Borrelly
488    and 21P/Giacobinni-Zinner.   Comparing their photometric age, P-AGE (13 cy, 14 cy, 20
489    cy, respectively), it is apparent that all spill-over comets are young objects.   The
490    photometric-age of 103P was found above to be 13±2 cy, in agreement with this
491    conclusion.   This comet should be observed after aphelion, to pinpoint the precise
492    location of the turn off, but this requires reaching to magnitude 26 and thus a large mirror.

493

494         After aphelion the comet is in a peculiar situation.  Its brightness is decreasing  in
495    spite of the fact that it is approaching the sun.

496

497    *1991 and 2004 apparitions*   To complete the study of this comet, we have compiled the
498    1991 and the 2004 secular light curves, defined with fewer observations.

499

500    In 1991 the envelope of the comet  was identical to that of 1997, thus the
501    physical parameters deduced are very similar to those of 1997 and are listed in Figure 6.
502

503    In 2004 there are only 25 observations (Figure 7).  However in this apparition the
504    comet perihelion jumped from 0.953 AU in 1991 to 1.036 AU in 2004.  There was then a
505    jump of 0.0833 AU.  The comet received less solar energy, and this is reflected in the
506    absolute magnitude m(1,LAG) that jumped from 7.4 in 1991 to 8.9 in 2004, a jump of 1.5
507    magnitudes.  This is the *perihelion effect* described in Paper VI and defined as: *"a change*
508    *in the absolute magnitude due to a jump in perihelion distance"*.      The parameters of this
509    apparition are listed in Figure 7.
510

511    *Future observational conditions*      This comet will have a close encounter with Earth on
512    October 21[st], 2010, at $\Delta$= 0.12 AU.  The object is in the observing list of the Arecibo radio
513    telescope (Harmon et al., 1999).  The encounter with the EPOXI spacecraft is scheduled
514    to take place on November 4[th], 2010, 24 days before perihelion.
515

516    **4. Water Budget**
517

518    We define the water budget, WB, as the total amount of water expended by the
519    comet in a single orbit.  Festou (1986) found a correlation between the water production
520    rate $Q_{H2O}$ and the reduced visual magnitude for many comets of the form
521

522        Log $Q_{H2O}$ = a – b m(1,1)                                    (4)
523

524    Subsequently many other researchers have found the same correlation (Roettger et
525    al., 1990; Jorda et al.,  2008; de Almeida et al., 1997).  In the latest determination by Jorda
526    et al. (2008) the following values for the parameters a and b were found  a = 30. 675,  b=
527    -0.2453                                    (5)
528

529    Using this empirical correlation it is possible to convert the reduced magnitudes
530    presented in the time plot (Figure 5) to water production rate, day by day.  The sum of
531    these values from $T_{ON}$ to $T_{OFF}$ gives the water budget in kg
532

533        $T_{OFF}$

534 $$WB = \sum_{T_{ON}} Q_{H2O}(t) \, \Delta t \tag{6}$$

535

536

537     In Figure 8 we plot the daily water production rate vs time to show how it
538 increases as a function of time. We also plot the same value for comet 9P/Tempel 1 for
539 comparison, since this comet was visited by a spacecraft and has been well studied
540 (A'Hearn et al. 2005, and papers therein). We conclude that 103P produces about 1.47
541 times more water than comet 9P in spite of being much smaller in diameter (1.43 vs 5.5
542 km).

543

544     Let us define a new age, the water budget age, thus

545

546     WB-AGE [cy] = $3.58 \times 10^{+11}$ / WB

547

548     The constant is chosen so that comet 28P/Neujmin 1 has a WB-AGE of 100 cy.
549 The WB-AGE is calculated for several comets in Table 2. We find that the WB-AGE = 19
550 cy for 103P. Compare this value with a WB-AGE = 28 cy for comet 9P/Tempel 1 once
551 again indicating that 103P is younger (more active) than 9P.

552

553     In Table 2 we compare the water production rates of several comets to place 103P
554 in perspective. We list the amplitude of the secular light curve, $A_{SEC}$, the photometric age
555 P-AGE(1,1), the photometric age P-AGE(1,q), and the new water budget age, WB-AGE.
556 Notice that WB spans a range of 5 orders of magnitude while P-AGE spans a range of
557 three orders of magnitude. Column 8 of Table 2 gives the WB scaled to 1P/Halley.

558

559     All the parameters of 103P have intermediate values, and thus we have to conclude
560 that we are dealing with a relatively young object. Notice for example that its P-AGE(1,1)
561 = 13 cy compares well with its WB-AGE = 19 cy. These values for comet 9P/Tempel 1
562 were P-AGE = 21 cy and WB-AGE = 28 cy. The WB scaled to comet 1P/Halley is also
563 of great interest. We find that comet 103P expended 3.4 % of the water of 1P, while 9P
564 expended 2.3 %.

565

566     **5. Validation of the concept of Photometric Age**

567

568    In Figure 9 the envelopes of several comets are compared taken from Paper VI.
569 We see that older comets are nested inside younger objects.   In other words, P-AGE
570 classifies comets by shape of the SLC.   Additionally Figure 9 shows that as a function
571 of age, $A_{SEC}$ and $R_{SUM}$, diminish in value.  That is, a comet has to get nearer to the sun
572 to get activated, and it is less and less active as it ages.

573

574    From Figure 9 and Table 2 the conclusion is that WB and P-AGE measure
575 activity of a comet and that activity diminishes monotonically as a function of age.

576

577 **6. Applicability of the concept of Af$\rho$**

578

579    The concept of Af$\rho$ was developed by A'Hearn et al. (1984) to quantify the dust in
580 a comet, using a radial model of constant velocity of expansion.    In this way
581 measurements carried out with different apertures, at different times and of different
582 comets, are directly comparable.  However, there are restrictions to the applicability of
583 this concept.  1) The model is not applicable within a few 100 km of the nucleus
584 because the dust is still being accelerated by the gas.  2) Beyond about $10^5$ km the
585 model also fails because solar radiation pressure affects the motion of the dust.   3) If
586 the comet has already turned off in its activity but a cloud of old dust moves with it, the
587 model is not right either.  4) Any comet with short term variability also fails to comply
588 with the Af$\rho$ concept. 5) Additionally, the model was not developed to deal with comet
589 tails or trails.  6) Since what is being measured is the dust, observations have to be
590 made in the red or infrared part of the spectrum.   Visual data, like the observations
591 presented in the secular light curves, measures mostly the gas represented by the $C_2$
592 molecular bands.

593

594    In general it is found that the Af$\rho$ model works best for a symmetric coma in
595 steady state, from $10^2$ to $10^5$  km approximately.  In this range it is expected that the
596 value will be constant with aperture.   If the dust (and thus the magnitude) follows a $1 / \rho$
597 distribution, then Af$\rho$ will be constant at every aperture size.  However notice that a $1 / \rho$
598 distribution gives an infinite integral for the total magnitude.  Thus *necessarily* the
599 intensity must decrease faster than $1 / \rho$ at large distances or the magnitude will not
600 converge.

601

602     Since the units of Af$\rho$ are cm, the question is what is the physical meaning of this

603     quantity.   In that regard it is useful to point out that Arpigny et al. (1986) have found a

604     correlation between Af$\rho$ and dust production in tons (A'Hearn, 2009, personal

605     communication).  They find that a value of Af$\rho$ of 1000 cm corresponds to a mass loss

606     rate of dust of roughly 1000 kg per second.    Alternatively an Af$\rho$ of 1 cm would

607     correspond to a mass loss rate of 1 kg per second.    This result ignores the size

608     distribution of particles among other parameters, but is a useful way to interpret the

609     meaning of Af$\rho$.  Unfortunately, the correlation has never been published.

610

611     The formula developed by A'Hearn et al. (1984) and also used by Meech et al.

612     (2009) is

613

614     $A(\theta)f\rho = [2.467 \ 10^{19} \ R^2 \ \Delta \ 10^{0.4(m_\odot - m)}] / \rho$                                    (7)

615

616     where $\theta$ = phase angle in degrees, R = Sun-comet distance in AU,   $\Delta$ = Earth-comet

617     distance in AU, $m_\odot$ = solar magnitude in the red = -27.1, m = observed magnitude, $\rho$ =

618     radius of aperture in arc seconds.   This equation implies that Af$\rho$ is a function of the

619     Sun-comet-Earth angle.   Schleicher et al. (1998) derived a quadratic fit to the comet

620     Halley data of the form

621

622

623     $\Delta \ Log \ A(\theta)f\rho = -0.01807 \ \theta + 0.000177 \ \theta^2$                                    (8)

624

625     that they use to model comet 9P/Tempel 1 data.   They mention that this phase function

626     works well for phase angles smaller than 45-50º.   The value in Equation (8) may be

627     multiplied by -2.5 to convert to magnitudes.

628

629     That the application of the Af$\rho$ concept is not straight forward can be plainly seen

630     comparing the results of Schleicher (2007) with those of Milani et al. (2007) for comet

631     9P/Tempel 1.  Figure 4c of Schleicher (2007) shows an Af$\rho$ peaking -25 days before

632     perihelion which is not in accord with the visual secular light curve of Ferrín (Paper III)

633     (Ferrin's Figure 2, which measures mainly the gas) peaking -10±5 days before

634    perihelion.    On the other hand Figure 4 of    Milani et al. (2007) shows Af$\rho$ peaking -

635    85 days before perihelion.    None of the results agree with the fact that the gas and dust

636    must be coupled on physical grounds.    These results would be very difficult to reconcile

637    with physical theories and would require fancy explanations.

638

639    In view of the limitations of the applicability of the Af$\rho$ concept, it is always wise to

640    check if the hypothesis of its constancy vs aperture is fulfilled.    Going back to our

641    measurements, in Figure 10a we see the magnitude of the comet in increasing

642    apertures, used to derive the infinite aperture magnitude (IAM) the night 050112.   It can

643    be seen that the IAM is reached for apertures larger than 42" = 73.000 km, much larger

644    that apertures typically used in cometary photometry.    That is why we emphasize the

645    need to use this curve of growth method to derive the total magnitude of the comet.    In

646    Figure 10b, we plot the Af$\rho$ value vs measuring aperture calculated with equation (7) to

647    see if the hypothesis of constant Af$\rho$ with $\rho$ is fulfilled.    We see that for the night of

648    050112 it is not.

649

650    An identical result is found for the night 060229.    An aperture of at least 10" =

651    31.000 km is needed to extract all the flux.    Once again we find that the conditions to

652    have a constant value are not fulfilled.    This is not surprising in view that the comet

653    exhibits a tail and thus solar radiation pressure is important, distorting the even flow of

654    the dust and making the conditions to have a constant Af$\rho$ invalid.

655

656    Schleicher (2007) finds a similar situation for comet 9P/Tempel 1.   He finds (his

657    Figure 3) that Af$\rho$ is strongly dependent on $\rho$ for five dates.   The question then arises, if

658    Af$\rho$ is a function of $\rho$, then what is the correct value of Af$\rho$?    Schleicher (2007) decides

659    to normalize all values to $10^4$ km from the nucleus and at a phase angle of 41º.    These

660    are arbitrary numbers that do not have attached any physical meaning.    We prefer to

661    correct to zero phase angle, and to identify the correct Af$\rho$ value as the one that

662    corresponds to the IAM derived in Figure 10a, and thus to the minimum value of Af$\rho$

663    corresponding to those apertures.    In this case the correct value of Af$\rho$ is smaller than

664    what we would derive using other criteria.

665

666       Our phase angles are given in Table    1. We find that $A(\theta)f\rho$ had a value of

667   around 55 cm on 050112 and correcting to $\theta = 0º$ with Equation (8), we find 106 cm.

668   For the date 060229 we find a $A(\theta)f\rho = 27$ cm, and when corrected to $\theta = 0º$ we find

669   $A(0)f\rho = 37$ cm.    These correspond to mass loss rates of 106 kg/s and 37 kg/s

670   respectively. The value decreases with distance to the Sun as expected theoretically.

671

672       In conclusion, in spite of the fact that the concept of $Af\rho$ is a useful one,

673   researchers have not always been aware of its limitations. The value of  $Af\rho$ must be

674   checked as a function of aperture to verify the applicability of the concept.    Thus

675   previous determinations of $Af\rho$ should probably be revised downward.

676

677   **7. Mass loss and remaining time**

678

679       From Table 2 it is possible to deduce that the comet lost $1.88 \times 10^{10}$ kg of water in

680   the 1997 apparition. However we are interested in the total mass loss. To calculate it

681   we need the dust to gas mass ratio, $\delta$.    The values of $Af\rho$ deduced in a previous

682   section are of no help because they are far from perihelion and are non-constraining.

683   Sykes and Walker (1992) favor a mean value $\delta = 2.9$ for a sample of several comets

684   while Singh et al. (1992) give values from 0.5 to 3. With this information it is possible to

685   calculate the thickness of the layer lost per apparition using the formula

686

687             $\Delta r = ( \delta + 1 ) \; WB / 4 \pi r^2 \rho$                          (9)

688

689   where r is the radius and  $\rho$ the density. This equation comes from the fact that the

690   density is given by $\rho = \Delta M/\Delta V$.    $\Delta V$, the volume removed, is given by $\Delta V = 4\pi r^2 \Delta r$. And

691   $\Delta M$, the mass removed, is given by $\Delta M_{H2O} + \Delta M_{DUST} = WB ( 1 + \Delta M_{DUST}/WB )$. For the

692   density we are going to take a value of $530$ kg/m$^3$ which is the mean of many

693   determinations compiled in Paper I. The resulting values of $\Delta r$ are compiled in Table 2

694   for 14 comets.

695

696       From Table 2 we see that comet 103P lost 34 m in radius in the 1997 apparition.

697   Since the radius of this comet is only 570 m, the ratio $r_N / \Delta r = 17$. This calculation implies

698   that the comet will sublimate away in only 17 additional revolutions, if the mass loss rate

699   is constant. This is the remaining time left for this comet, and it is a very small time (~109

700    y).  However this is dependent on the dust    to    gas    mass    ratio    assumed.    Future
701    observations should be acquired to follow the rapid sublimation of this comet and to
702    pinpoint more precisely its remaining time.
703
704    **8. Comets Visited by Spacecraft**
705
706         From Table 2 and Figure 11 we find that by chance all comets that have been
707    visited by spacecraft are young: 1P/Halley (7 cy), 9P/Tempel 1 (21 cy), 19P/Borrelly (14
708    cy), 81P/Wild 2 (13 cy).  103P/Hartley 2 follows the same trend with P=AGE = 13 cy.
709    Thus we expect the surface of 103P to resemble that of previous comets.
710    67P/Churyumov-Gerasimenko will be visited in 2014 by Rosetta.  This comet has P-
711    AGE = 32 cy (Paper I) and thus it should present a slightly older surface.
712
713         What would be interesting to see is the surface of a really old object, a methuselah
714    comet (P-AGE > 100 cy).  Several are available.  The following photometric ages have
715    been extracted from the *Atlas of Secular Light Curves* (Paper VI) and unpublished results:
716    107P/Wilson-Harrington (760 cy), 133P/Elst-Pizarro (280 cy), 169P/NEAT = 2002 EX12
717    (245 cy), P/2006 T1 Levy (100 cy), 162P/Siding-Spring (230 cy).  Not so old comets but
718    still methuselah are 2P/Encke (102 cy) and 28P/Neujmin 1 (100 cy).   A spacecraft
719    mission to any methuselah object would be of great scientific interest.   It is specifically
720    predicted that a Methuselah comet will have a more extreme surface morphology, than the
721    cometary surfaces that have been imaged up to now.
722
723    **9. Conclusions and results**
724         The main conclusions of this work are:
725
726    a)  We have shown that for comets that exhibit a coma, the best estimate of the brightness
727    is the infinite aperture magnitude, IAM, obtained using a curve of growth method (Figure
728    2).  This is usually 0.4 - 2.0 magnitudes brighter than magnitudes deduced with smaller
729    apertures (Figures 1, 2, Section 2).
730
731    b) Of the order of 30 photometric parameters are measured and listed in the plots of this
732    comet, over 20 of them new (Figures 4 to 7).   The turn on point of activity is -4.2±0.1 AU
733    from the Sun, which corresponds to -400±40 d before perihelion.  The total active time is $T_{ACTIVE}$ = 1904 ±43 d.  For comparison the  active time of comets 1P/Halley and 9P/Tempel 1 are $T_{ACTIVE}$ = 1992 and 1069 days.   The comet exhibits a lag between the maximum brightness and perihelion, LAG= +16±2 days that may be due to a thermal lag or to the pole pointing to the sun at that time, or a combination of both (Figures 4 and 5).

c) In Paper VI comets are classified according to their age and diameter.  It is proposed there that comets with 4 < P-AGE < 30 cy are young, while comets with D < 1.5 km are dwarf objects.  In Section 3 we find the photometric age P-AGE= 13±2 cy, while the diameter found from the literature is 1.14±0.16 km.   Thus according to the above classification 103P is a *young dwarf comet.*  For comparison comets 1P/Halley and 9P/Tempel 1 have photometric ages P-AGE= 7.1 cy and P-AGE= 22 cy (Paper VI) and diameters 9.8 km and 5.5 km.

d) The nucleus is *very active* as can be deduced from the fact that the amplitude of the SLC is $A_{SEC}$= 10.8±0.1 mag in 1997.   For comparison comets 1P and 9P have $A_{SEC}$ = 10.8 and 9.0 mag.  From a sample of 27 comets (Paper VI), it is apparent that there is a maximum value of  $A_{SEC}$ = 11.55±0.1 mag.  103P is close to that maximum value, implying that we are seeing a relatively young object (Figures 4 and 5).

e) The validity of the concept of photometric age, P-AGE, is studied comparing the envelopes of several comets.  It is apparent that older comets are nested inside the envelope of younger comets.  The conclusion is that P-AGE classifies comets by shape of their secular light curve (Figure 9).   This plot also shows that as a function of age, the amplitude of the secular light curve, $A_{SEC}$, and $R_{SUM}$ = $-R_{ON}+R_{OFF}$, decrease in value. The comet has to get nearer to the sun to get activated and its activity is less and less (Section 5).

f) This comet belongs to the class of *spill-over* comets defined in Paper VI.   After remaining active up to aphelion, the comet spills-over its activity into the next orbit.  The other comets that populate this class are also young (Figure 4 and 5).

g) 103P also belongs to the family of *comets that exhibit a break point* in their SLC.  Two linear laws change slope at a break point located at $R_{BP}$ = -1.20±0.1 AU before perihelion, and magnitude $m_{BP}$ = 9.4±0.1.  The slope of the first linear law after turn on is n= +9.44.

768 This value can be compared with the slopes of comets Hale-Bopp, 1P/Halley and
769 9P/Tempel 1: +10.3, +8.92 and +7.7 respectively. Since the SLC after turn on does not
770 exhibit curvature in any of these comets, it is safe to conclude that sublimation is not
771 controlled by water ice. The controlling substance may be CO or $CO_2$ ice. $CO_2$ has been
772 detected spectroscopically.

773

774 h) The water budget of this comet is calculated. The comet expends $1.88 \ 10^{10}$ kg of water
775 per apparition vs $1.28 \ 10^{10}$ kg for comet 9P/Tempel 1. A new water-budget age is defined
776 and it is found that WB-AGE = 19 cy vs WB-AGE = 28 cy for 9P. These results imply that
777 this is a comet younger than 9P/Tempel 1.

778

779 i) The values of Afρ are calculated on two dates and we find dust production rates of 106
780 and 37 kg/s, decreasing with solar distance as expected. The applicability of the concept
781 of Afρ is studied using a double plot of magnitude and Afρ vs aperture. In this way infinite
782 aperture magnitudes and infinite aperture Afρs are deduced (Figure 10 and Section 6).

783

784 j) The thickness of the layer lost per apparition is calculated, and we find Δr = 34 m where
785 r is the radius. Since the radius of this comet is r = 570 m, it is found that at the present
786 rate the comet may disappear in only 17 revolutions or ~109 y. However this is
787 dependent on the dust to gas mass ratio assumed which is uncertain.

788

789 k) By chance all comets visited by spacecraft have small photometric ages and thus are
790 young objects (Figure 11 and Section 8). 103P follows the same trend. Thus it is
791 expected that the surface morphology that will be found in future EPOXI images should be
792 very similar to that of previous comets. It is suggested that any new mission to a comet
793 should be made to a really old object and several are proposed (107P, 133P, 162P, 169P,
794 and P/2006 T1).

795

796 The *"Atlas of Secular Light Curves of Comets"* in color as well as numerous tables
797 of cometary physical properties are available and can be downloaded from the web site:
798 http://webdelprofesor.ula.ve/ciencias/ferrin.

799

800

801

802 **10. Acknowledgements**

803       To two unknown referees for having contributed significantly to the scientific
804 improvement of this paper. To the Council for Scientific, Technologic and Humanistic
805 Development of the University of the Andes for their support through grant number C-
806 1281-04-05-B. To the Centro de Investigaciones de Astronomía, CIDA, for the time
807 granted to observe comets. The help of the night assistants and professional observers,
808 Ubaldo Sanchez, Freddy Moreno, Orlando Contreras, Gregory Rojas, Daniel Cardozo,
809 Faviola Moreno, Cecilia Mateu, and Carlos Castillo, is highly appreciated. We thank the
810 members of the comet-obs web site for providing the partial aperture magnitudes: I.
811 Almendros, F. Baldris, J. Barceló, M. Camarasa, M. Campas, M. Casao, E. Cortés, J.
812 Castellanos, CEAMIG-REA, J. A. de los Reyes, J. L. Dorestes, F. Fratev, F. García, T. C.
813 García, J. García M., J. Lacruz, L. and S. Lahuerta, D. Mendicini, F. Montalbán, R.
814 Morales, R. Naves, S. Pastor, E. Reina, M. Reszelsky, D. Rodríguez, J. L. Salto, J.
815 Sanchez, J. R. Vidal.
816
817

**11. References**

A'Hearn, M., Schleicher, D. G., Millis, R.L., Feldman, P.D., Thompson, D.T., 1984.
Comet Bowell 1980b. AJ, 89, 579-591.

A'Hearn, M.F. and 32 colleagues, 2005. Deep Impact: Escavating Comet Tempel 1.
Science, 310, 258-264

A`Hearn, M.F., Combi, M.R., 2007. Deep Impact at Comet Tempel 1. Icarus,187, 1-3.

Arpigny, C., Dossin, F., Woszczyk, A., Donn, B., Rahe, J., Wyckoff, S., 1986.
Presentation of an Atlas of cometary spectra. In Asteroids, comets, meteors II;
Proceedings of the International Meeting, Uppsala, Sweden, June 3-6, 1985
(A87-11901 02-90). Uppsala, Sweden, Astronomiska Observatoriet.

Chen, J., Jewitt, D., 1994. On the Rate at Which Comets Split. Icarus, 108, 265-271.

Colangeli, L., Epifani, E., Brucato, J.R., Bussoletti, E., de Sanctis, C., Fulle, M., Mennella,
V., Palomba, E., Palumbo, P., Rotundi, A. 1999. Infrared spectral observations of
comet 103P/Hartley 2 by ISOPHOT. A&A, 343, L87-L90.

Crovisier, J., Encrenaz, Th., Lellouch, E., Bockelee-Morvan, D., Altieri, B., Leech, K.,
Salama, A., Griffin, M.J., de Graauw, Th., van Dishoeck, E.F., Knacke, R., Brooke,
T.Y. 1999. ISO Observations of short period comets. In "The Universe seen by
ISO". P. Cox and M.F. Kessler (Eds.), ESA SP427, 137-140.

Crovisier, J., Brooke, T.Y., Leech, K., Bockelee-Morvan, D., Lellouch, E., Hanner, M.S.,
Altieri, B., Keller, H.U., Lim, T., Encrenaz, Salama, Griffien, M., de Graauw, T., van
Dishoeck, E., Knacke, R.F. 2000. The thermal infrared spectra of comets
Hale-Bopp and 103P/Hartley 2 oberved with the infrared Space Observatory.
In Thermal Emission Spectroscopy and Analysis of Dust, Disks, and
Regoliths. M.L. Sitko, Al.L. Spraegue and D.K. Lynch (Eds.). ASP Conference
Series, 196, 109-117.

De Almeida, A.A., Singh, D.D., Hubner, W.F., 1997. Water release rates, active areas,
and minimum nuclear radius derived from visual magnitudes of comets-an
application to comet 46P/Wirtanen. PSS, 45, 681-692.

Delsemme, A.H., 1983. Chemical Composition of Cometary Nuclei. P. 92, in Comets,
L.L. Wilkening, Editor. Univ. Of Arizona Press, Tucson, AZ.

Ferrín, I., 2005a. Secular Light Curve of Comet 28P/Neujmin 1, and of Comets Targets
of Spacecraft, 1P/Halley, 9P/Tempel 1, 19P/Borrelly, 21P/Grigg-Skejellerup,
26P/Giacobinni-Zinner, 67P/Chruyumov-Gersimenko, 81P/Wild 2. Icarus
178, 493-516. Paper I.

Ferrín, I., 2005b. Variable Aperture Correction Method in Cometary Photometry,
ICQ 27, p. 249-255. Paper II.

Ferrín, I., 2006. Secular Light Curve of Comets, II: 133P/Elst-Pizarro, an asteroidal
belt comet. Icarus, 185, 523-543. Paper III.

Ferrín, I., 2007. Secular Light Curve of Comet 9P/Tempel 1. Icarus, 187, 326-331.
Paper IV.

Ferrín, I., 2008. Secular Light Curve of Comets 2P/Encke, a comet active
at aphelion. Icarus, 197, 169-182. Paper V.

Ferrín, I., 2010. Atlas of Secular Light Curves of Comets. PSS, 58, 365-391.
is available in color at http://webdelprofesor.ula.ve/ciencias/ferrin. Paper VI.

Festou, M.C., 1986. The derivation of OH gas production rates from visual magnitudes
Of comets. In Asteroids, Comets, Meteors II, C.I. Lagerkvist et al., Editors,
299-303. Uppsala Univ. Press, Uppsala, Sweden.

Fink, U., Hicks, M.P., Fevig, R.A., 1999. Production rates for the Stardust Mission Target:
81P Wild 2. Icarus, 141, 331-340.

Green, D.W.E., 1997. ICQ Guide to observing comets. Cambridge, MA; Smithsonian
Astrophysical Observatory, p. 100-103.

Green, D.W., 2006. International Comet        Quarterly, 28, 81.

Groussin, O., Lamy, P., Jorda, L., Toth, I., 2004.  The Nuclei of Comets 126P/IRAS
        and 103P/Hartley 2.  A&A, 419, p. 375-383.

Harmon, J. K., Campbell, D. B., Ostro, S.J., Nolan, M.C., 1999.  Radar observations
        of comets, Planetary and Space  Science, 47, p. 1409-1422.

Jorda, L., Crovisier, J., Green, D.W.E., 2008.  The correlation between water production
        rates and visual magnitudes of comets.  In Asteroids, Comets, Meteors 1991,
        285-288.  Lunar and Planetary Institute, LPI Contribution No. 1405, paper id
        8046, Houston, TX, USA.

Jorda, L., Lamy, P., Groussin, O., Toth, I., A'Hearn, M. F., Peschke, S., 2000.  ISOCAM
        Observations of cometary nuclei.  In Proceedings of ISO Beyond Point Sources,
        Studies of Extended Infrared Emission, (R. J. Lureijs et al., Editors), p. 61,
        ESA SP-455, Noordwijk, The Neatherlands.

Licandro, J., Tancredi, G., Lindgren, M., Rickman, H., Hutton, R. G., 2000.  CCD
        Photometry of Cometary Nuclei, I: Observations from 1990-1995.  Icarus,
        147, p. 161-179.

Lisse, C.M., Fernandez, Y.R., Reach, W.T., Bauer, J.M., A'Hearn, M.F., Farnham, T.L.,
        Groussin, O., Belton, M.J., Meech, K.J., Snodgrass, C.D., 2009.  PASP, 121,
        968-975.

Lowry, S.C., Fitzsimmons, A., 2001.  CCD Photometry of Distant Comets, II.  A&A,
        365, p. 204-213.

Lowry, S.C., Fitzsimmons, A., Collander-Brown, S., 2003.  CCD Photometry of Distant
        Comets, III.  A&A, 397, p. 329-343.

Milani, G.A., Szabo, Gy.M., Sostero, G., Trabatti, R., Ligustri, R., Nicolini, M., Facchini,
        M., Tirelli, D., Carosati, D., Vinante, C., Higgins, D., 2007.  Photometry of Comet
        9P/Tempel 1 during the 2004/2005 approach and the Deep Impact module
        impact.  Icarus, 187, 276-284.

Meech, K.J., Pittichova, J., Bar-Nun, A., Notesco, G., Laufer, D., Hainaut, O.R., Lowry,
        S.C., Yeomans, D.K., Pitts, M. 2009.  Activity of comets at large heliocentric
        distances pre-perihelion. Icarus 201, 719-739.

Roettger, E.E., Feldman, P.D., A'Hearn, M.F., Festou, M.C., 1990.  Comparison of water
        production rates from UV spectroscopy and visual magnitudes for some recent
        comets.  Icarus, 86, 100-114.

Schleicher, D., G., Millis, R.L., Birch, P.V., 1998. Narrowband Photometry of Comet
        P/Halley: Variation with Heliocentric Distance, Season, and Solar Phase Angle.
        Icarus, 132, 397-417.

Schleicher, D., G., 2007. Deep Impact's target Comet 9P/Tempel 1 at multiple
        apparitions: Seasonal and secular variations in gas and dust production.  Icarus,
        190, 406-422.

Singh, P.D., de Almeida, A.A., Huebner, W.F., 1992.  Dust release rates and dust to
        Gas mass ratios of eight comets.  An.J., 104, 848-858.

Sosa, A., Fernandez, J.A., 2009.  Cometary masses derived from non-gravitational
        forces.  MNRAS, 393, 192-214.

Snodgrass, C., Lowry, S.C., Fitzsimmons, A., 2006.  Photometry of cometary nuclei:
        Rotation rates, colours and a comparison with Kuiper Belt Objects.  MNRAS,
        385, 737-756.

Snodgrass, C., Meech, K.J., Hainaut, O.R., 2010.  The nucleus of 103P/Hartley 2,
        target of  The EPOXI mission.  A&A, May 10, 2010, in press.

Sykes, M.V., Walker, R.G., 1992.  The nature of comet nuclei.  Asteroids, Comets,
        Meteors, 1991, 587-591.

Weaver, H.A., Feldman, P.D., McPhate, J.B., A'Hearn, M.F., Arpigny, C., Smith,

T.E., 1994. Detection of CO Cameron band emission in comet P/Hartley 2 (1991 XV) with the Hubble Space Telescope. Ap.J., 422, 374-380.

**Figure captions**

Figure 1. Comet 103P/Hartley 2 imaged with the 1 m Schmidt telescope of the National Observatory of Venezuela, at f/3, on 2005, January 12[th]. Since large photometric apertures have to be used to extract the whole flux, the image has previously been cleaned of nearby stars using a cloning tool. The area between the two circles measures the sky background. Both images have the same scale. Left: A normal stretching of the image shows that an aperture of radius 20 pixels is apparently enough to extract a total magnitude. Stretching refers to the maximum and minimum pixel intensity to display the image in the computer monitor. The standard stretching is selected by the computer, who does not know about preserving the flux. Thus the monitor displays a deceivingly faint image. Right: A forced stretching reveals a much larger coma than expected and a much brighter comet. At least a 44 pixel radius is needed to extract a total magnitude (some flux is still left out). This has been called the *insufficient CCD aperture error* (Paper II), and manifests itself by producing measurements that lie much below the envelope of the observations.

Figure 2. Derivation of an infinite aperture magnitude. It can be seen how the calibration star and the comet increase in flux with increasing photometric apertures. The asymptotic value is the infinite aperture magnitude, IAM. Notice the huge aperture needed to extract the whole flux of the comet, much larger than apertures typically found in the literature. Notice also how the calibration star #14 converges rapidly to its asymptotic value but still needs an 8 pixel radius to retrieve an IAM.

Figure 3. Images of comet 103P/Hartley 2 taken with telescopes of the National Observatory of Venezuela. Upper row, 1m Schmidt telescope working at f/3. Bottom row, 1 m Reflector Telescope, at f/21. The information related to these images is compiled in Table 1. North is at the top, East to the left. The two left images are simultaneous observations. They are particularly interesting to compare because they were taken with very different f/ ratios. The f/21 reflector looses about 0.4 mag. Even in the upper right hand image taken at R= 5.03 AU from the sun and magnitude R($\Delta$,R)= 20.2, the comet displays a faint coma. $V_{IAM}$ and $R_{IAM}$ are infinite aperture magnitudes obtained by fitting an exponential decay function to magnitudes measured as a function of increasing CCD apertures (see text).

*Upper Left hand image*: 1m Schmidt, f/3. Date = 2005 01 12.19, Exposure = 21 min, $R_{IAM}(\Delta,R)$ = 17.1±0.1, $\Delta t$ = +134 d, R = 2.85 AU, $\Delta$= 2.40 AU, RA = 07h 24m 05s, DEC = +18º 58' 28", Filter = Clear. The image is 3.5'x5.0' in size.

*Upper Right hand image*: 1m Schmidt, f/3. Date = 2006 03 01.20, Exposure = 12 min, $R_{IAM}(\Delta,R)$ = 20.4±0.2, $\Delta t$ = +135 d, R = 5.03 AU, $\Delta$ = 4.31 AU, RA = 07h 24m 18s, DEC = +18º 55' 58", Filter = Clear. The image is 3.5'x5.0' in size

*Lower Left hand image*: 1m Reflector, f/21. Date = 2005 01 12.17, Exposure = 45 min, $R_{IAM}(\Delta,R)$ = 17.0±0.2, $\Delta t$ = +134 d, R = 2.85 AU, $\Delta$= 2.40 AU, RA = 07h 24m 05s, DEC = +18º 58' 28", Filter = Clear. The image is 3.5'x5.0' in size.

*Lower Right hand image*: 1m Reflector, f/21. Date = 2005 01 14.23, Exposure = 50 min, $R_{IAM}(\Delta,R)$ = 17.8±0.2, $\Delta t$ = +135 d, R = 2.87 AU, $\Delta$ = 2.39 AU, RA = 07h 24m 18s, DEC = +18º 55' 58", Filter = Clear. The image is 3.5'x5.0' in size

Figure 4. Secular light curve of comet 103P/Hartley 2 in 1997, log plot. The slope 5 line at the bottom of the plot in the form of a pyramid is due to the atmosphereless nucleus. The comet exhibits a well defined SLC, with a linear behavior before turn on indicating that it is controlled by a substance more volatile that water ice, probably $CO_2$ or CO. This is in agreement with the small photometric age, P-AGE= 13±2 comet years. The activity spills over into the next orbit as shown at right, where the brightness is ~5 magnitudes above the nucleus. Notice the cluster of modern CCD observations. The error of the Snodgrass et al. observation (SMH) is contained inside the size of the symbol. TW= this work. LF= Lowry and Fitzsimmons (2001). CJ= Chen and Jewitt (1994). LTLRH= Licandro et al. LFCW= Lowry et al. (1999), SLF= Snodgrass et al. (2006), SMH= Snodgrass et al. (2010), GLJT= Groussin et al. (2004). L et al. = Lisse et al. (2009).

Figure 5. Secular light curve of comet 103/Hartley 2 in 1997, time plot. The comet exhibits a prominent belly as expected of a young object. There is a cluster of modern CCD observations on the right hand side. The error of the Snodgrass et al. observation (SMH) is contained inside the size of the symbol. Figure 5 is a text-book example of a "spill over comet" with a prominent belly.

Figure 6. Secular light curve of comet 103P/Hartley 2 in 1991, log plot. The comet exhibits a very small perihelion effect (q-effect), defined as a change in the perihelion magnitude m(1,LAG) as a result of a change in the perihelion distance, q. See text.

Figure 7. Secular light curve of comet 103P/Hartley 2 in 2004, log plot. From 1997 to 2004 there was a change in m(1,1) of +1.5 magnitudes.

Figure 8. Water budget of 103P compared to the water budget of 9P/Tempel 1. Most of the sublimation takes place near perihelion, explaining the large slope at q. Although the comet is active at aphelion, the contribution to WB is negligible at that distance. The production rate of 103P is asymmetric with respect to perihelion with an asymmetry parameter of 1.76, in agreement with the asymmetry shown in Figure 5 in the time plot.

Figure 9. Validation of the concept of photometric age, P-AGE: envelopes of comets compared. It is apparent that older comets are nested inside the envelope of younger comets. The conclusion is that P-AGE classifies comets by shape of their secular light curve. On the left hand side, the comet number. On the right hand side, the photometric age. This plot also shows that as a function of age, the amplitude of the secular light curve, $A_{SEC}$, and $R_{SUM} = -R_{ON}+R_{OFF}$, decrease in value. The comet has to get nearer to the sun to get activated, and is less and less active.

Figure 10. Afρ of comet 103P. The Afρ is calculated as a function of aperture from Figure 2 and equation 7. Above the infinite aperture magnitude is calculated. Below it can be seen that Afρ does not stay constant with ρ, decreasing in value with larger apertures. The reason is that the coma was not spherically symmetric, and exhibited a tail. The Afρ model assumes a symmetric coma, not distorted by the solar wind pressure. What then is the correct value of Afρ? We believe it is the value derived when the whole flux has been extracted, and this corresponds to the infinite aperture magnitude.

Figure 11. By chance, all comets visited by spacecraft are young. Comet 103P/Hartley 2, with a photometric age of 13 comet years, follows the same trend. Thus we do not have any idea of how an old cometary surface looks like. We would expect to find a surface

morphology for 103P not very different from that of previously visited objects.   A mission to a really old comet would be scientifically very exciting.   Cometary targets to old comets, are suggested in the text (Section 9).

Table 1. Observing Log of comet 103P/Hartley 2.

| YYMMDD | Δt [d] | Diameter=1m Type    f/ | Δ [AU] | R [AU] | α [º] | Expo [s] | $R_{IAM}(Δ,R)$ | $V_{IAM}(Δ,R)$ | $V_{IAM}(1,R)$ |
|--------|--------|------------------------|--------|--------|-------|----------|-------------|-------------|-------------|
| 050112 | +212   | Schmidt  f/3           | 2.40   | 2.85   | 19.7  | 1260     | 16.7±0.1    | 17.2±0.1    | 15.3±0.1    |
| 050112 | +212   | Reflector f/21         | 2.40   | 2.85   | 19.7  | 2700     | 17.0±0.1    | 17.5±0.1    | 15.6±0.1    |
| 050114 | +214   | Reflector f/21         | 2.39   | 2.87   | 19.4  | 3000     | 17.8±0.1    | 18.3±0.1    | 16.4±0.1    |
| 060301 | +625   | Schmidt  f/3           | 4.31   | 5.03   | 8.5   | 700      | 20.2±0.2    | 20.7±0.2    | 17.5±0.2    |
| 070216 | +1004  | Schmidt  f/3           | 4.92   | 5.49   | 9.7   | 834      | >20.9±0.3   | >21.3       | >17.6       |
| 070220 | +1008  | Schmidt  f/3           | 4.92   | 5.78   | 9.6   | 1807     | >21.3±0.4   | >21.7       | >18.0       |

YMD= Year, Month, Date. Δt [d] = time after perihelion. Δ [AU] = Earth-comet distance. R [AU] = Sun-comet distance. α [º] = phase angle. Expo [s] = Total Exposure Time. $R_{IAM}(Δ,R)$ = infinite aperture R-band magnitude at Δ, R. $V_{IAM}(Δ,R)$ = equivalent V-band magnitude obtained using <V-R>= 0.41±0.16 for comet 103P (Paper II). $V_{IAM}(1,R)$= V-band magnitude reduced to Δ= 1 AU.

Table 2. Comparison of 103P/Hartley 2 with other comets, in order of increasing water budget age, WB-AGE.
WB-AGE [cy] = 3.58 E+11/ WB [kg]  and P-AGE [cy] = 1440 / [ $A_{SEC} \cdot R_{SUM}$ ] ; $R_{SUM}$ = $-R_{ON} + R_{OFF}$

| Comet | WB [kg] | $R_{SUM}$ [AU] | $A_{SEC}$ (1,1) | P-AGE (1,1) [cy] | P-AGE (1,LAG) [cy] | WB-AGE (1,q) cy | WB ---------% WB(1P) | $r_N$ [km] | $Δr_N$ [m] | $r_N$ ----- $Δr_N$ |
|-------|---------|------------|-----------|-----------|-------------|------------|--------------|---------|---------|---------|
| HB    | 7.66 E+12 | 52.1 | 11.5 | 2.3  | 2.3   | 0.047 | 1393  | 27    | 6.2    | 4388   |
| 1P    | 5.52 E+11 | 18.7 | 10.8 | 7.1  | 6.6   | 0.65  | 100   | 4.9   | 13     | 364    |
| Hya   | 2.46 E+11 | 6.7  | 11.6 | 19   | 17    | 1.5   | 44.6  | 2.4   | 25     | 96     |
| 109P  | 1.29 E+11 | 5.9  | 8.2  | 29.7 | 28    | 2.8   | 23.4% | 13.5  | 0.4    | 32571  |
| 65P   | 3.06 E+10 | 13.0 | 10.8 | 10   | 13    | 12    | 5.5%  | 3.7   | 1.3    | 2826   |
| 81P   | 2.09 E+10 | 10.2 | 11.4 | 13   | 17    | 17    | 3.8%  | 1.97  | 3.2    | 624    |
| **103P** | **1.88 E+10** | **9.9** | **10.7** | **15** | **14** | **19** | **3.4%** | **0.57** | **33.9** | **17** |
| 2P    | 1.33 E+10 | 3.1  | 4.8  | 98   | 64    | 27    | 2.4%  | 2.55  | 1.1    | 2321   |
| 9P    | 1.27 E+10 | 7.7  | 9.0  | 21   | 26    | 28    | 2.3%  | 2.75  | 0.98   | 2796   |
| 45P   | 7.95 E+09 | 3.5  | 7.5  | 55   | 40    | 45    | 1.4%  | 0.43  | 25.2   | 17     |
| 28P   | 3.58 E+09 | 4.5  | 3.2  | 100  | 133   | 100   | 0.6%  | 11.5  | 0.016  | 725892 |
| 26P   | 1.64 E+09 | 3.2  | 5.2  | 85   | 89    | 218   | 0.3%  | 1.47  | 0.44   | 3307   |
| 133P  | 1.81 E+08 | 0.2  | 1.4  | ---- | 280*  | 1978  | 0.32% | 2.3   | 0.020  | 114796 |
| 107P  | 9.68 E+07 | 0.3  | 3.3  | ---- | 760*  | 3698  | 0.17% | 1.65  | 0.021  | 79249  |

*For comets 107P and 133P the definition of P-AGE(1,1) fails, so we give T-AGE(1,q) instead.

1152
1153
1154
1155
1156
1157
1158
1159
1160
1161
1162
1163
1164
1165
1166
1167
1168
1169
1170
1171
1172
1173
1174
1175

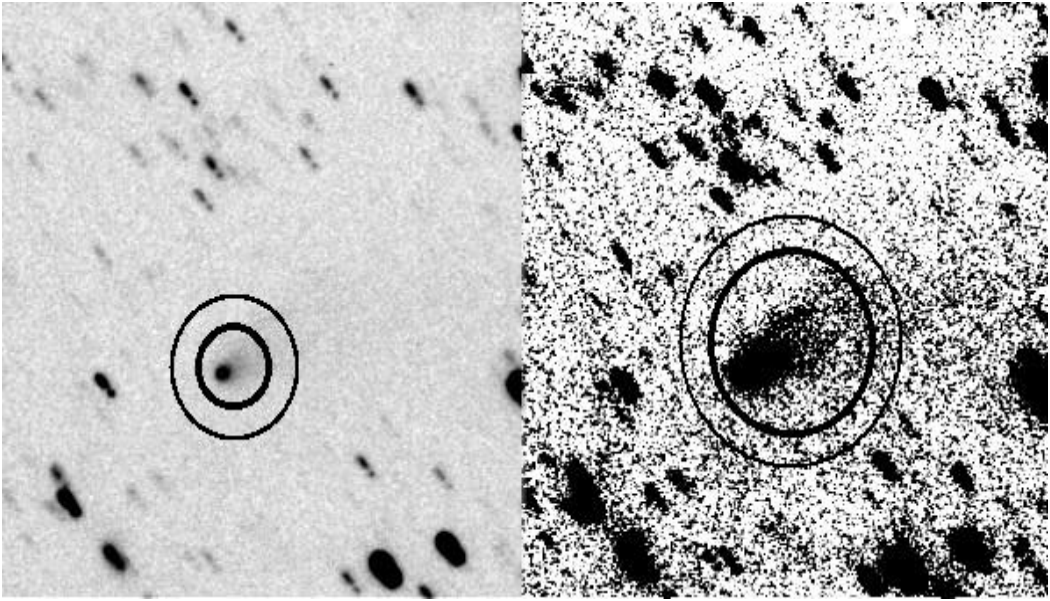

Figure 1.

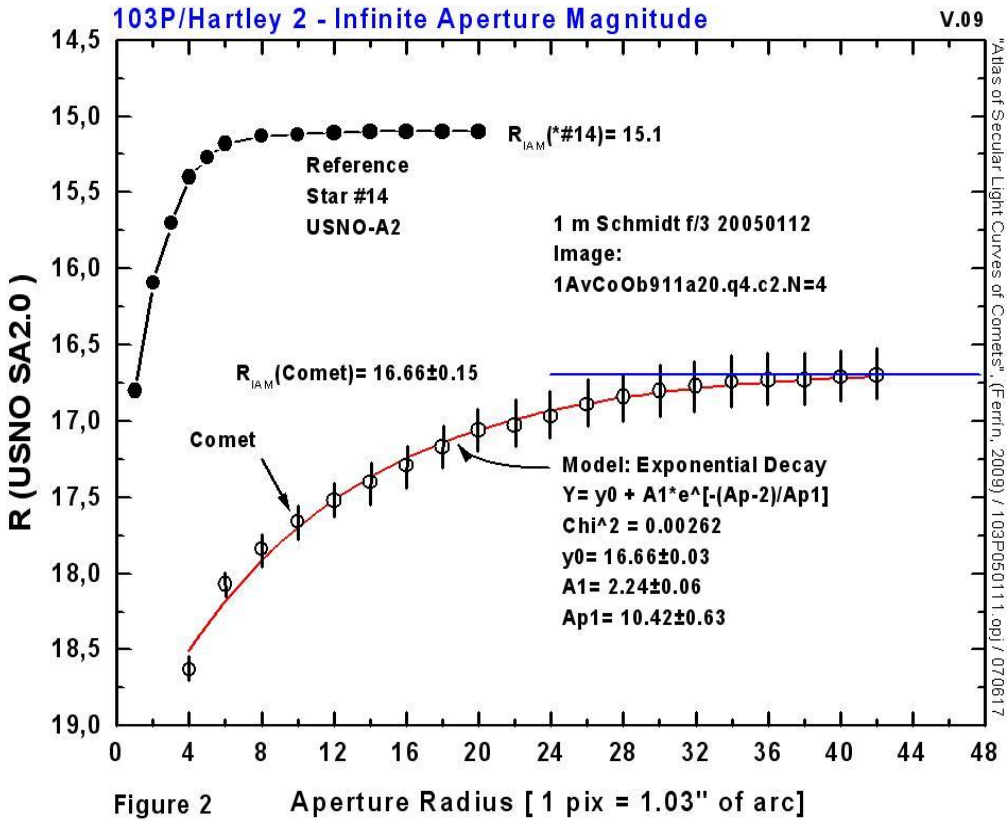

Figure 2

1176
1177

1178
1179

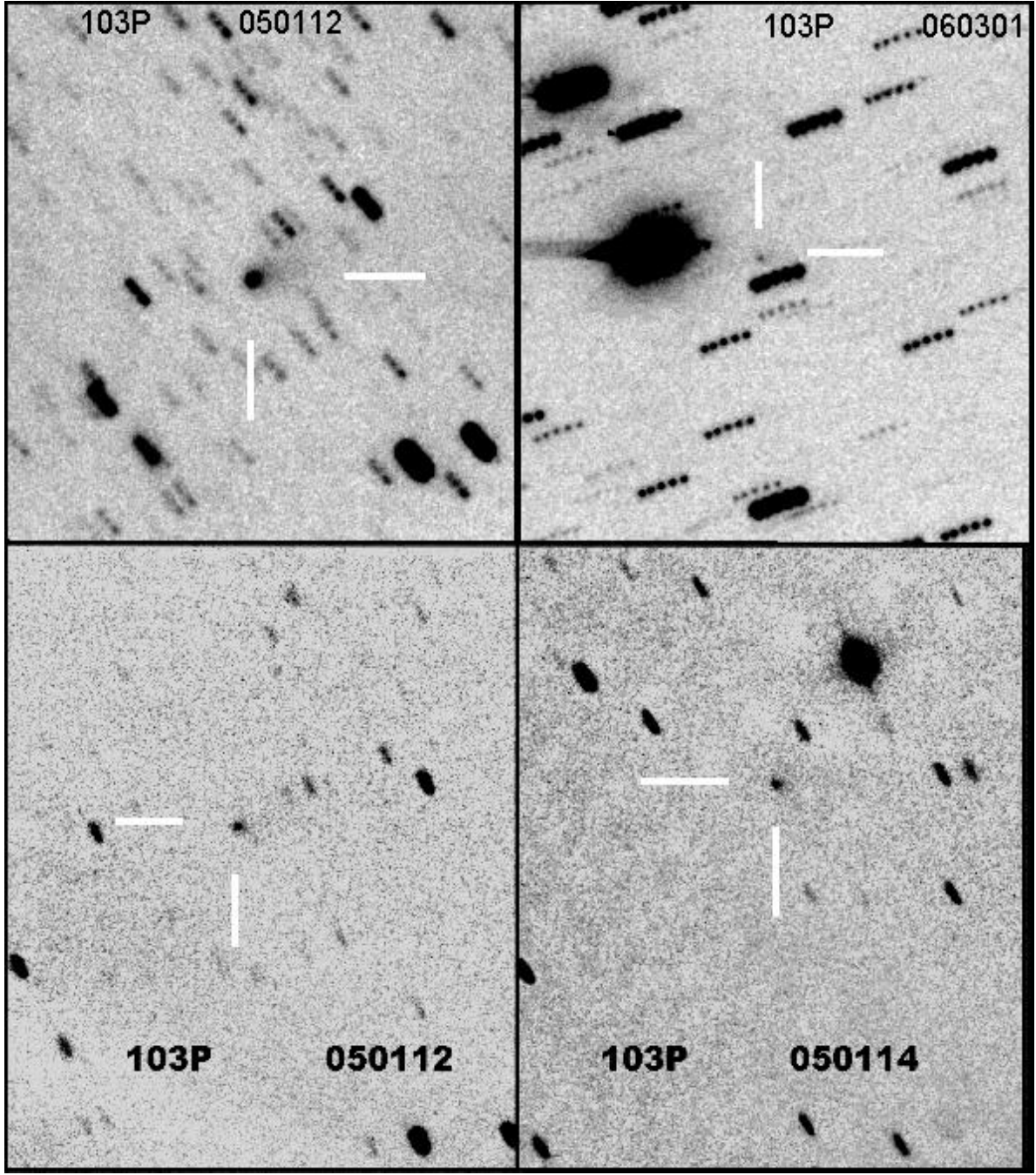

1180
1181
1182
1183
1184
1185
1186
1187
1188
1189
1190
1191
1192

Figure 3

1193
1194
1195

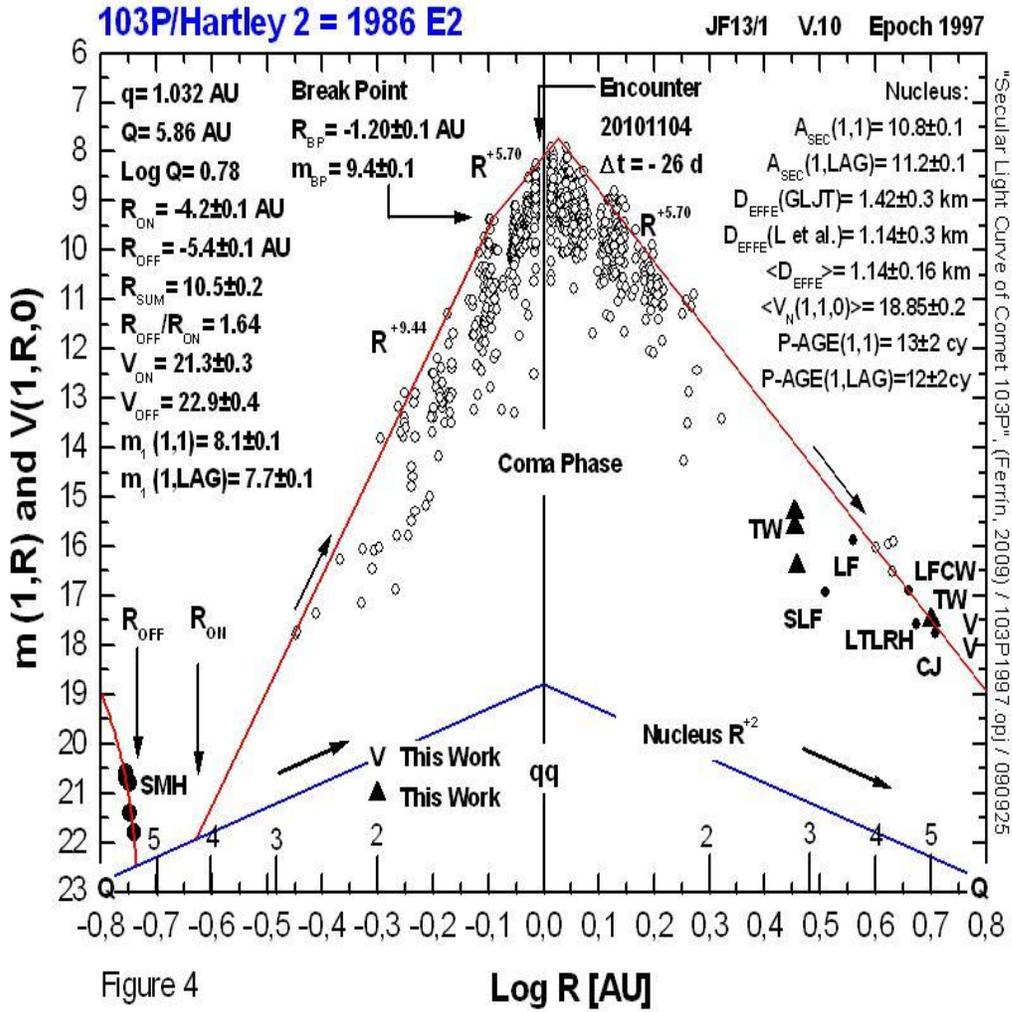

Figure 4

1196
1197
1198
1199
1200
1201
1202
1203
1204
1205
1206
1207
1208
1209
1210

Figure 4

1211
1212
1213
1214

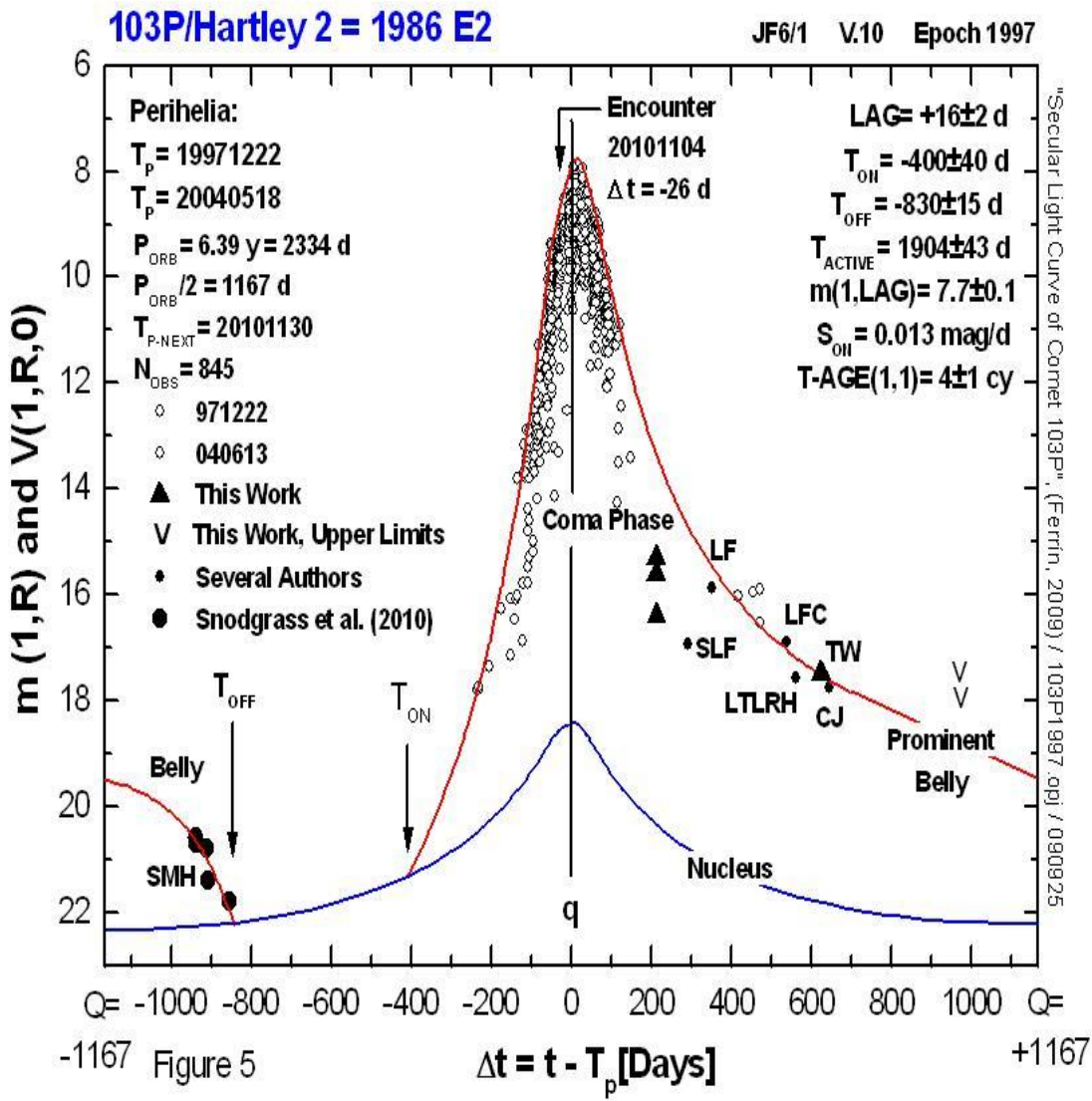

Figure 5

1215
1216
1217
1218
1219
1220
1221
1222
1223
1224
1225
1226
1227

1228
1229
1230

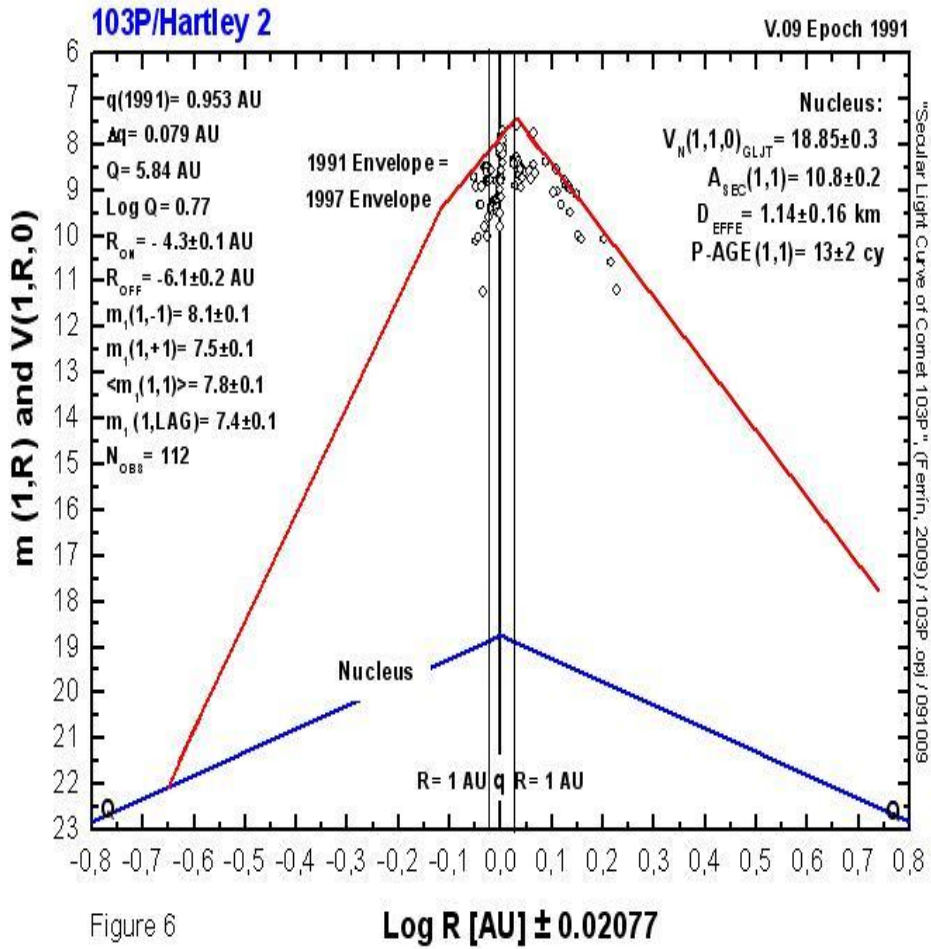

1231
1232
1233
1234                                    Figure 6
1235
1236
1237
1238
1239
1240
1241
1242
1243
1244
1245
1246
1247
1248

1249
1250
1251
1252

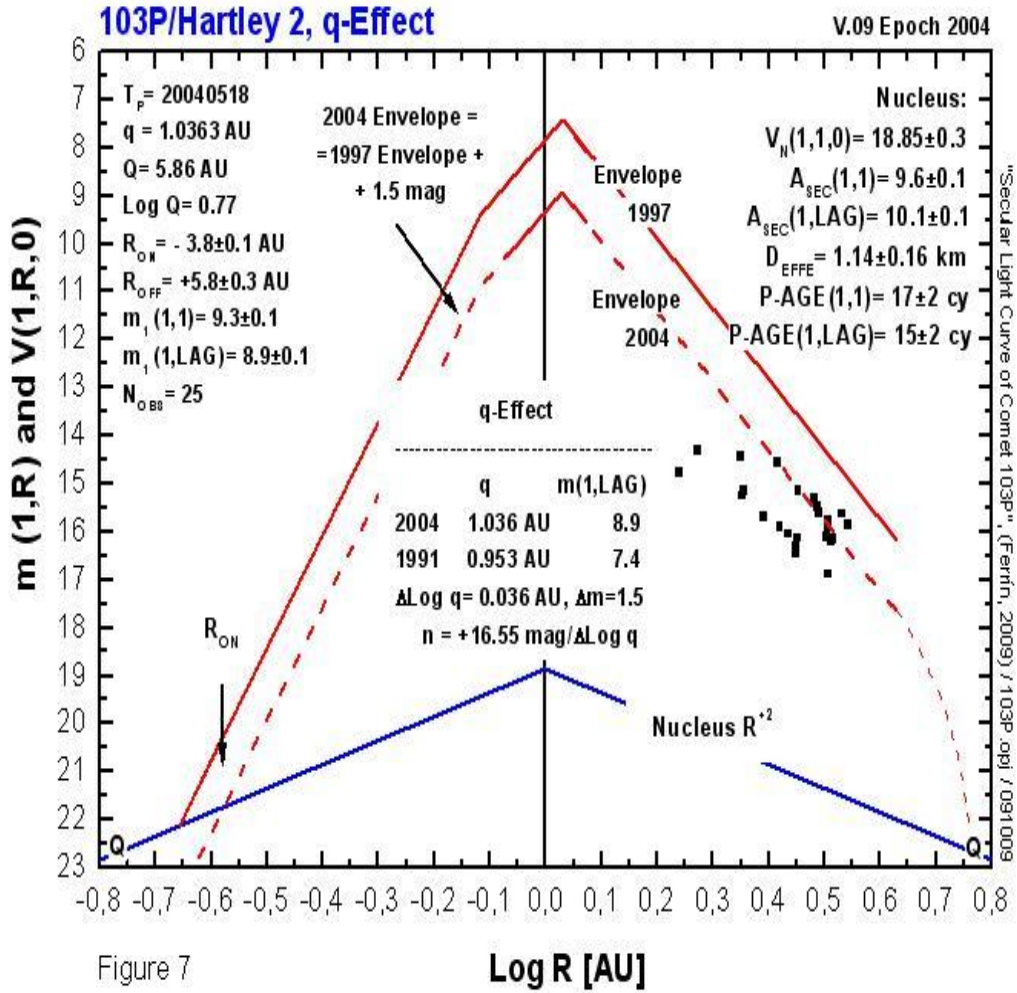

Figure 7

1253
1254            Figure 7
1255
1256
1257
1258
1259
1260
1261
1262
1263
1264
1265
1266
1267
1268

1269
1270

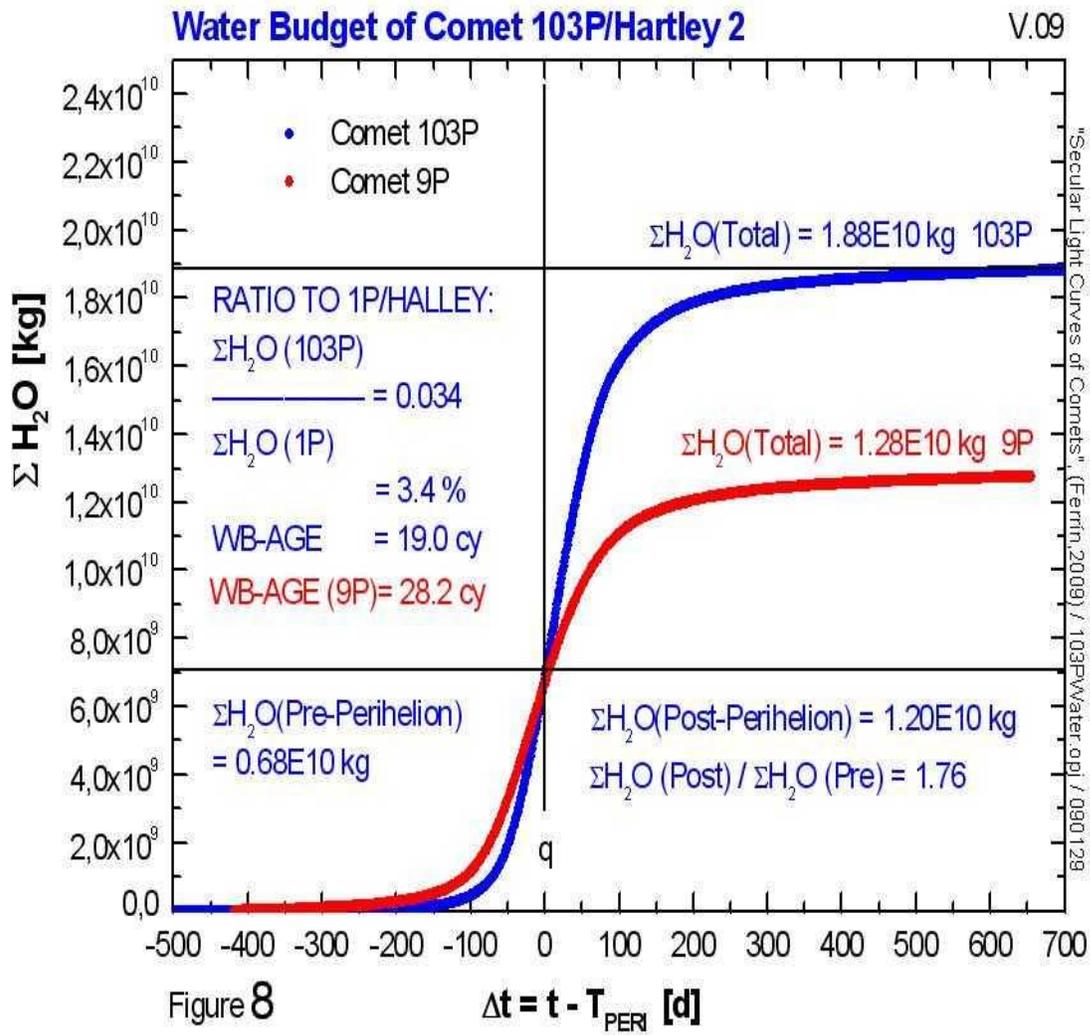

Figure 8

1271
1272
1273
1274
1275
1276
1277
1278
1279
1280
1281
1282
1283
1284
1285
1286
1287

Figure 8

1288
1289

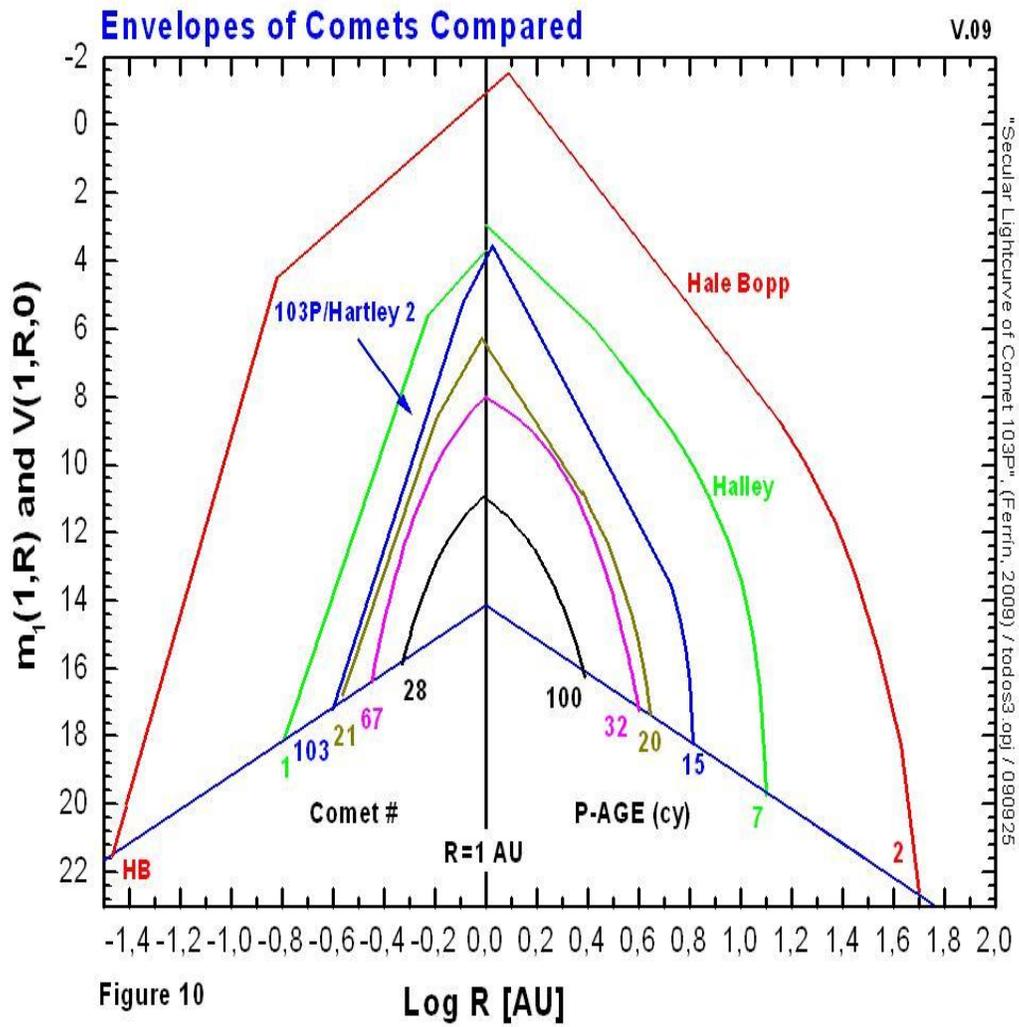

Figure 10

1290
1291
1292                                              Figure 9
1293
1294
1295
1296
1297
1298
1299
1300
1301
1302
1303
1304

1305
1306

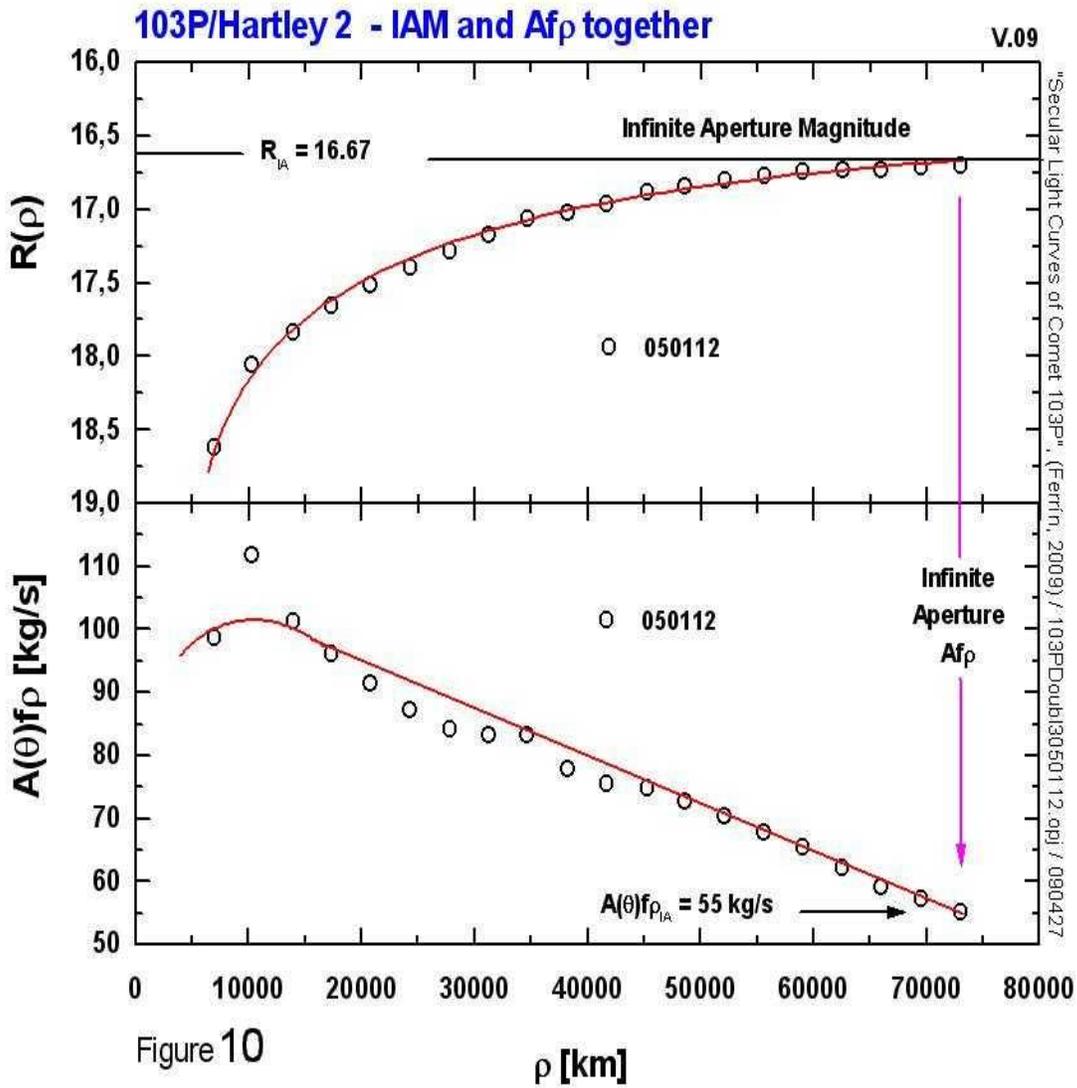

Figure 10

1307
1308
1309 Figure 10
1310
1311
1312
1313
1314
1315
1316
1317
1318
1319
1320
1321
1322
1323

1324
1325

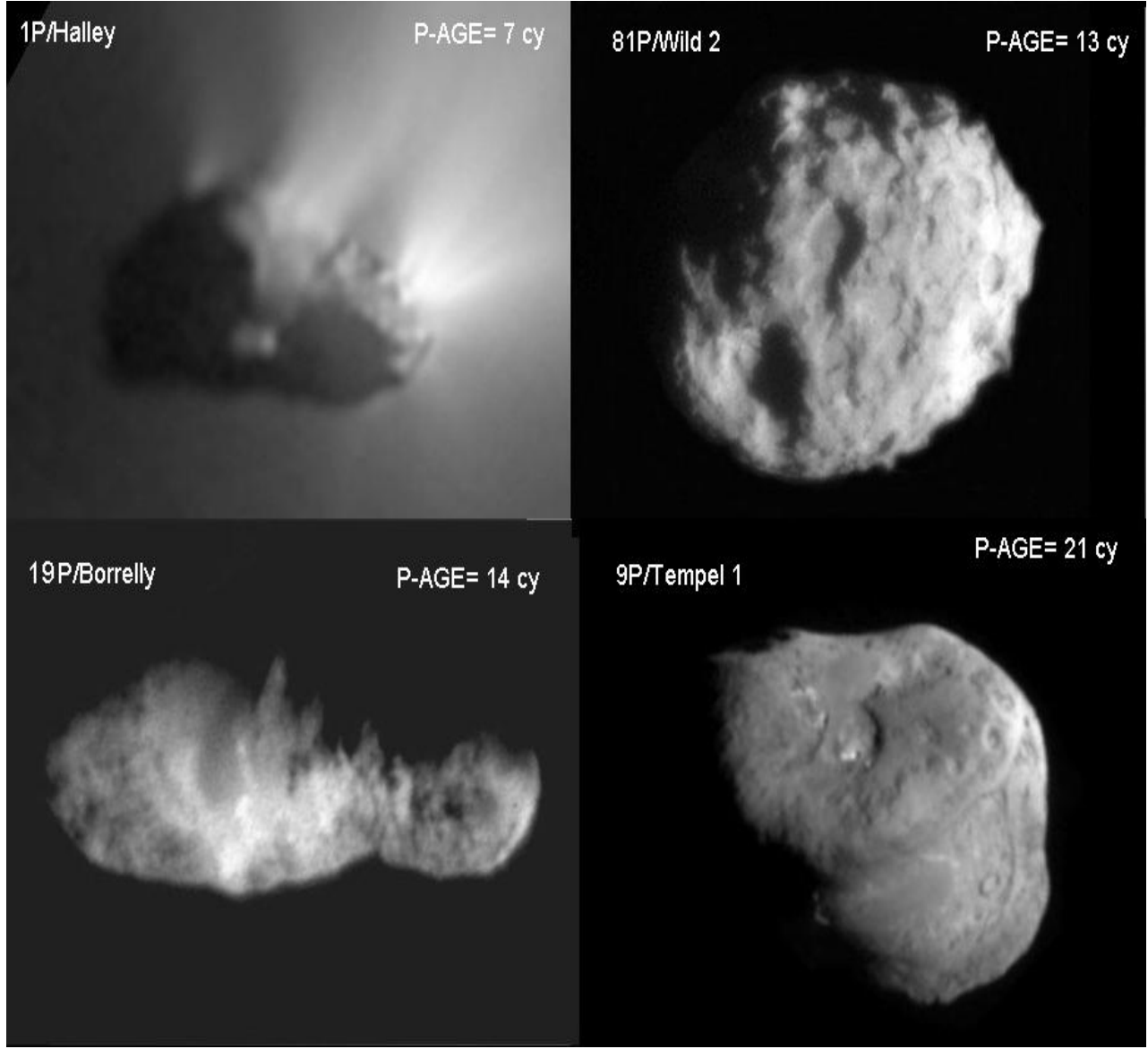

1326
1327
1328
1329
1330
1331          Figure 11
1332
1333
1334
1335
1336
1337
1338
1339
1340
1341
1342